\documentclass[12pt]{iopjournal}

\usepackage{graphicx}
\usepackage{dcolumn}
\usepackage{bm}
\usepackage{multirow}
\usepackage{amsmath,amssymb,amsfonts}
\usepackage[dvipsnames]{xcolor}

\usepackage{ragged2e}
\AtBeginDocument{\justifying}

\begin{document}

\articletype{Paper}

\title{Accelerating high-fidelity gyrokinetic fusion simulations with iGENE}

\author{Gabriele Merlo$^1$,  Victor Artigues$^1$, Carl-Martin Pfeiler$^1$, and Frank Jenko $^1$}

\affil{$^1$Max Planck Institute for Plasma Physics, Garching Munich, DE}

\begin{abstract}
We present the latest version of iGENE, a differentiable implementation of the local (flux-tube) gyrokinetic code GENE, now fully rewritten in JAX. iGENE features a multispecies, collisional, fully electromagnetic nonlinear gyrokinetic model while preserving end-to-end differentiability. The code is verified against GENE for linear ITG, TEM and KBM physics and for nonlinear turbulent fluxes with adiabatic and kinetic electrons. Forward-mode derivatives of the linear growth rate agree with finite differences for all five parameters tested, resolving the 30--50\% discrepancy in $\hat{s}$ and $\epsilon$ reported for the earlier TensorFlow version. All collectives are native JAX operations, and the code strong-scales from 2 to 32 H100 GPUs with 86\% parallel efficiency. We then assess three routes to a shorter time to solution. Custom CUDA FFT kernels operating in bfloat16 are up to $3.97\times$ faster than fp32 cuFFT on the transforms, and on a single H100 iGENE is $2.63\times$ faster per time step than single-precision GENE. Multi-rate sub-cycling of the parallel dynamics yields a further $1.89\times$ speedup with kinetic electrons. Finally, a Hermite--Laguerre velocity-space representation is implemented, offering tunable fidelity.
\end{abstract}

\section{Introduction}
Fusion energy is one of the most formidable scientific and engineering challenges, among others because of the complex, nonlinear, multi-scale, multi-physics phenomena that determine the behavior of a confined plasma. Simulation tools remain critical for fusion research and for probing physics that experiments cannot currently access. Specific numerical methods have been and are continuously developed to produce more and better data. An obvious question is thus how to accelerate such simulations, often requiring extreme computational effort, while maintaining the necessary degree of fidelity for the downstream tasks.

This work reports our efforts to accelerate high fidelity local gyrokinetic simulations, which are the workhorse for characterizing turbulent transport in magnetically confined plasmas. Following the proof of principle discussed in \cite{Artigues_2026}, we have further developed the iGENE code, adding all the missing physics elements that were originally not available in the TensorFlow implementation and are necessary for a high fidelity description of a fusion plasma. The code now implements a fully electromagnetic collisional model, is interfaced to most common tokamak and stellarator equilibrium solvers and has been verified against the Fortran version of GENE~\cite{jenko2000electron}. The code has been fully rewritten in JAX~\cite{jax2018github}, keeping end-to-end automatic differentiability as a crucial property. It can therefore be used in optimization workflows as well as AI/ML pipelines.  

We have verified the code results and measured its parallel performance, which has improved from the TensorFlow port. The iGENE algorithm, a faithful translation of the GENE one, is already extremely optimized on massively parallel architectures, but it remains computationally expensive. We have therefore explored different paths to reduce the time to solution. In particular, modifications to the time stepping scheme enabling sub-cycling of the parallel dynamics, the usage of mixed precision arithmetic and a spectral description of velocity space. The latter is a new feature added to iGENE motivated by the success of similar representations including e.g. the GX code~\cite{Mandell_2024} or GENE-X~\cite{Frei_2025}. It is targeted specifically to a multi fidelity approach to turbulence, where the number of moments can be selected to reduce the cost of a run. Furthermore, thanks to the use of agentic AI, the code has been strongly optimized for GPU usage by writing custom FFT kernels that outperform vendor library, an effort that was made easier by adopting JAX. Thanks to these efforts iGENE is now faster than the original GENE code, by $2.63\times$ per time step on a single H100 for the kinetic-electron case of Sec.~\ref{sec:mixed_precision} when compared with GENE in single precision, although certain optimization strategies could be backported to Fortran and equivalently applied there.

The remainder of the paper is as follows. Section \ref{sec:implementation} presents the numerical model and details its discretization. Section \ref{sec:validation} presents an extensive code verification, both of direct quantities of interest and their gradients and Sec.~\ref{sec:performance} its performance. Section~\ref{sec:acceleration} discusses the implementation of accelerating techniques, including time stepping, mixed precision and velocity space representation. An outlook is provided in Section~\ref{sec:conclusion}.

\section{The numerical model and its numerical discretization}
\label{sec:implementation}
We have extended the numerical model of iGENE to include collisional and fully electromagnetic effects i.e., including magnetic fluctuations parallel and perpendicular to the background magnetic field. The equations are well known and are reported here for the sake of completeness and to illustrate the features now available in iGENE. For a general derivation of the nonlinear gyrokinetic theory we refer to e.g. Ref.\cite{Brizard_2007}.

The dynamics of a multispecies plasma are obtained by solving the time evolution of the gyrocenter distribution $F_s({\bf X},v_{\|},\mu,t)$ for each plasma species $s$ according to the gyrokinetic Vlasov equation:
\begin{equation}
\frac{\partial F_s}{\partial t}+\dot{\mathbf{X}}\cdot\nabla
F_s+\dot{v}_{\parallel}\frac{\partial F_s}{\partial v_\|}=C[F_s]\,,
\label{eq0}
\end{equation}
with ${\bf X}$ the gyrocenter position, $\mu$ the magnetic moment and $v_{\|}$ the velocity component parallel to the background magnetic field $\mathbf{B}={\rm B}\mathbf{b}$. $C[F_s]$ is the collision operator. More details will be provided in the following. The equations of motion of a gyrocenter with mass $m$ and charge $q$ read
\begin{align}
\label{eq1}
\dot{\mathbf{X}}&=v_{\parallel}\mathbf{b}+\frac{B}{B^*_{\parallel}}\left(\mathbf{v}_{\nabla  B}+\mathbf{v}_{\kappa}+\mathbf{v}_{\bf E}\right), \\
\label{eq2}
\dot{v}_{\parallel}&=-\frac{\dot{\mathbf{X}}}{mv_{\parallel}}\cdot\left(\mu(\nabla B +
\nabla\bar{B}_{\parallel})+q\nabla\bar{\phi}\right) -\frac{q}{m}\dot{\bar{A}}_\parallel
\end{align}
where $\mathbf{v}_{\nabla B}=(\mu/(m\Omega B))\,\mathbf{B}\times\nabla B$ is the grad-B drift velocity, $\mathbf{v}_{\kappa}=(v_{\parallel}^2/\Omega)\,\left(\nabla\times\mathbf{b}\right)_{\perp}$
is the curvature drift velocity, and $\mathbf{v}_{\bf E}=(1/B^2)\,\mathbf{B}\times\nabla(\bar\phi-v_\|\bar{A}_\|+\frac{\mu}{q}\bar{B}_{\parallel})$ is the generalized $\mathbf{E}\times\mathbf{B}$ drift velocity. Here, $\Omega=qB/m$ is the gyrofrequency and $B^*_{\parallel}$ the parallel component of the effective magnetic field $\mathbf{B}^*=\mathbf{B}+\frac{B}{\Omega}v_{\parallel}\nabla\times \mathbf{b}+\nabla\times\left(\mathbf{b}\bar{A}_{\parallel}\right)$. 
Finally, $\bar\phi$ and $\bar{A}_\parallel$  are the gyroaveraged electrostatic potential and  parallel component of the vector potential. $\bar B_\parallel$ is the gyroaveraged parallel magnetic-field fluctuation.

In solving Eq.~\ref{eq0} we use the delta-f method, splitting the distribution function $F$ in a fixed background $F_0$ and a fluctuating part $f$.  Currently only local Maxwellian backgrounds are supported. Akin to GENE's electromagnetic scheme, iGENE introduces the modified distribution function 
\begin{equation}
g = f - (q/m)\,\bar A_\parallel\,\partial{F_0}/\partial{v_\parallel}
\label{eq:g_to_f}
\end{equation}
so that $\dot{\bar A}_\parallel$ in Eq.\ref{eq2} can be absorbed into a single time derivative and $g$ is the state evolved in time.\footnote{This scheme corresponds to the linearized limit of the mixed-variable formulation of Mishchenko \textit{et al.}~\cite{Mishchenko}.}

The system is closed self-consistently by the field equations, obtained from Maxwell's equations through a linearized gyrocenter-to-particle pullback. In the local limit the fields are represented in Fourier space in $x$ and $y$, so that $\nabla_\perp^2\to-k_\perp^2$, with $k_\perp^2=g^{xx}k_x^2+2g^{xy}k_xk_y+g^{yy}k_y^2$ given by the metric coefficients of the field-aligned coordinates, and gyroaverages reduce to multiplications by Bessel functions, e.g.\ $\bar\phi=J_0(\lambda_s)\phi$ with $\lambda_s=k_\perp v_\perp/\Omega_s$ and $v_\perp=\sqrt{2\mu B/m_s}$.

Since the Debye length is small compared to the scales of interest, Poisson's equation reduces to the quasineutrality condition $\sum_s q_s n_s=0$, where the particle density $n_s$ obtained from the pullback consists of the gyroaveraged gyrocenter density and the polarization density. Neglecting $B_\parallel$ for clarity, it reads
\begin{equation}
\sum_s q_s\int J_0(\lambda_s)\, f_s\, d^3v
  = \sum_s \frac{q_s^2 n_{0s}}{T_{0s}}\left[1-\Gamma_0(b_s)\right]\phi ,
\label{eq:qn}
\end{equation}
with $d^3v=(2\pi B/m_s)\,dv_\parallel d\mu$, $\Gamma_0(b)=I_0(b)e^{-b}$ and
$b_s=k_\perp^2 T_{0s}/(m_s\Omega_s^2)$. When electrons are treated adiabatically they contribute a term $(q_s^2n_{0s}/T_{0s})(\phi-\langle\phi\rangle)$ to the right-hand side,
where $\langle\phi\rangle$ is the flux-surface average, nonzero only for $k_y=0$.
Since $g-f$ is odd in $v_\parallel$, $f$ can be replaced by $g$ in
Eq.~\eqref{eq:qn}.

The parallel vector potential follows from the parallel component of Amp\`ere's law (neglecting the displacement current). Expressed in terms of $g$, it reads
\begin{equation}
\left[k_\perp^2+\mu_0\sum_s\frac{q_s^2n_{0s}}{m_s}\Gamma_0(b_s)\right]A_\parallel
  = \mu_0\sum_s q_s\int v_\parallel J_0(\lambda_s)\, g_s\, d^3v ,
\label{eq:ampere}
\end{equation}
where the second term in brackets (the skin term) arises from the change of variables $f\to g$.

Finally, in the gyrokinetic ordering the perpendicular component of
Amp\`ere's law reduces to perpendicular pressure balance,
\begin{equation}
\frac{B\,B_\parallel}{\mu_0} = -\sum_s \delta p_{\perp,s} ,
\label{eq:bpar}
\end{equation}
where $\delta p_{\perp,s}$ is the perturbed perpendicular pressure in particle space. Its gyrocenter contribution involves the moment $\int \mu B\,\frac{2J_1(\lambda_s)}{\lambda_s} f_s\, d^3v$, while the pullback introduces additional terms in $\phi$ and $B_\parallel$. As a result, Eqs.~\eqref{eq:qn} and \eqref{eq:bpar} form a coupled  linear system for $(\phi, B_\parallel)$ at each $(k_x,k_y,z)$, which is solved
directly; see Refs.~\cite{merz2008gyrokinetic, Goerler_2011} for the explicit coefficients.
An arbitrary number of species can be simulated, including adiabatic ions or electrons. Electromagnetic fields can be switched on/off via $\beta$.


iGENE currently solves Eq.\eqref{eq0} in the local (flux-tube) limit~\cite{beer1995field}, the extension to nonlocal simulations is left for future work. We use the same numerical discretization of GENE, which on the one hand has proven to be particularly effective and on the other allows us to quickly debug and verify the code correctness. The distribution function is discretized on a fixed structured grid in a five-dimensional phase space, which consists of three configuration (labeled $x, y, z$) and  two velocity space $v_\parallel$ and $\mu$ directions. Spatial coordinates are field-aligned. The underlying axisymmetry of a tokamak  corresponds to an invariance of the unperturbed system with respect to the toroidal angle, which translates to an invariance with respect to $y$. Consequently, fluctuating fields related to a linear eigenmode correspond to a single $k_y$ Fourier mode. The local limit allows the usage of periodic boundary conditions also in the radial directions, thus both $x$ and $y$ directions are implemented in Fourier space using pseudospectral methods. The parallel direction is instead discretized with fourth-order finite differences. Standard twist-and-shift \cite{beer1995field} as well as the generalized version given in \cite{Martin_2018} are supported. An equidistant grid is used in $v_\parallel$ while a Gauss-Laguerre quadrature is used in $\mu$. An alternative spectral discretization in velocity space using Hermite-Laguerre polynomials as basis function has been introduced in the code as an option to accelerate simulations at a lower fidelity. We refer the reader to Section~\ref{sec:acceleration_mom}. Default time integration is explicit using a standard fourth-order Runge-Kutta (RK4) scheme with an adaptive time step and a first-order splitting of collisions via a Runge-Kutta-Chebyshev scheme~\cite{Sommeijer_1998}. Again, faster alternatives have been explored as discussed in Section~\ref{sec:subcycling}. The $E\times B$ nonlinearity is evaluated pseudospectrally with dealiasing (3/2 rule). Numerical dissipation in the form of fourth order hyperdiffusion in $z$, $v_\parallel$, and optionally in $x,y$ is available to the user, including the adaptive LES model in \cite{Navarro_2014}. Finally, external  $E\times B$ sheared flows are supported as discussed in \cite{McMillan_2019}.

Different collision operators have been implemented in iGENE, equivalent to the ones available in the original Fortran code version. While GENE provides a very extensive suite of operators, we have limited iGENE to the ones that are de-facto used for production runs: linearized Landau-Boltzmann and Sugama. Both are implemented with the finite-volume schemes of Ref.~\cite{Crandall_2020}, which by construction conserve particles, momentum and energy and satisfy Boltzmann's H-theorem; we refer the interested reader to that work and references therein for a detailed description.

Several geometric interfaces are available in iGENE, both analytical as well as numerical. They cover both tokamak and stellarator use cases. Once again, we favored the interfaces that are most commonly used and allow for complete differentiability across the entire calculation. Hence iGENE implements s-alpha, circular~\cite{lapillonne2009clarifications}, Miller~\cite{Candy_2009}, EFIT ~\cite{Lao_1985} (currently non differentiable but widely used in the tokamak community and valuable for benchmarking against GENE), VMEC and GVEC~\cite{gvec_joss_2026} (also non differentiable at this stage).

Finally, although possible with limited effort, iGENE currently does not support pure eigenvalue calculations, with the exception of the evaluation of the limiting linear CFL condition \cite{Doerk_2014}.

Particular care was taken in keeping inputs and outputs of iGENE compatible
with GENE (both in terms of normalizations as well as file format) so that a
user can easily transition from one code to the other and benchmarking is
significantly simplified. The same normalizations as GENE are used in the
numerical implementation \cite{merz2008gyrokinetic}.

\section{Code verification}
\label{sec:validation}
Given that we have completely rewritten the code in JAX and added additional features compared to what discussed in \cite{Artigues_2026}, we first perform a complete verification of the code in the different regimes it can be used.

\subsection{Linear physics}
\begin{figure*}[ht]
    \centering
    \includegraphics[width=0.75\textwidth]{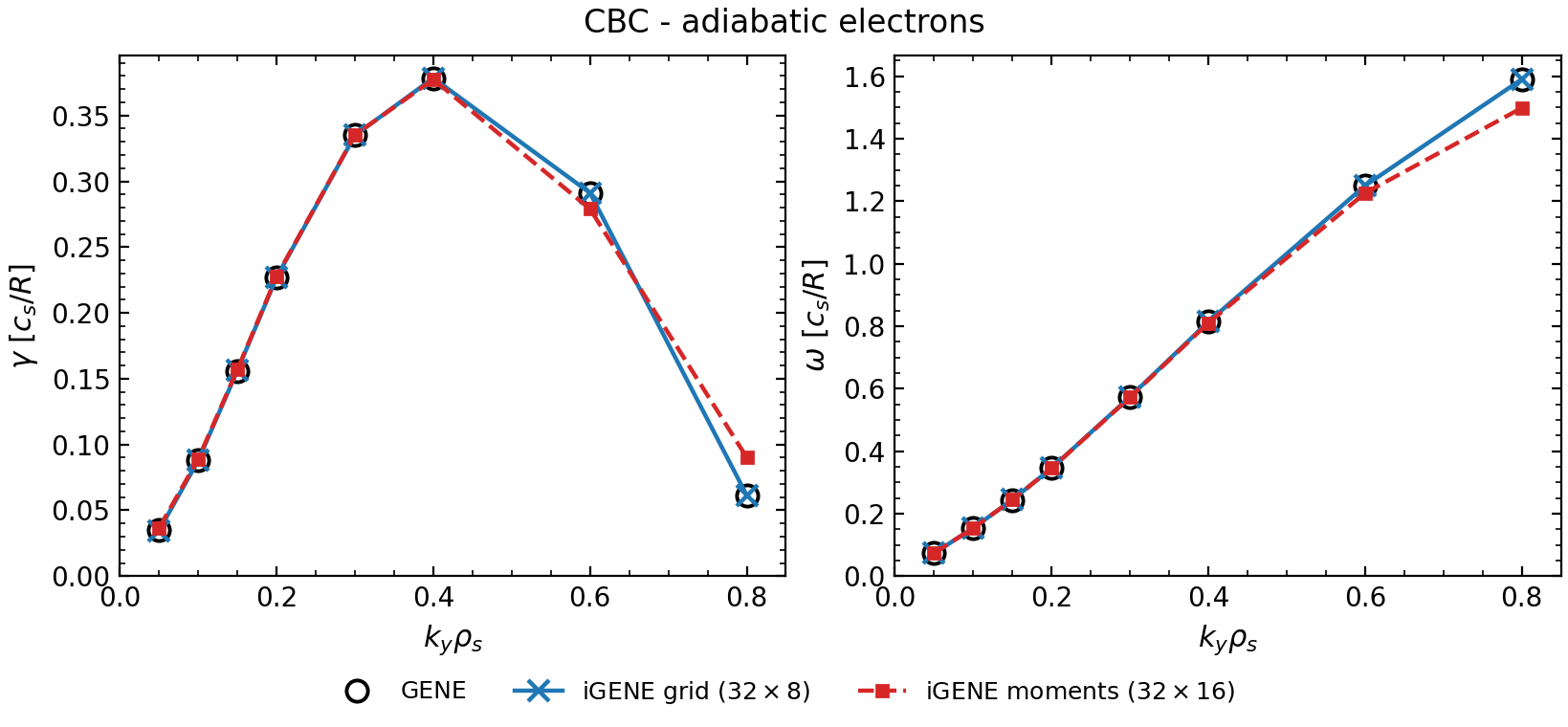}
    \caption{(left) Growth rate $\gamma$ and (right) frequency $\omega$ in units of $c_s/R$ for the CBC case with adiabatic electrons. Shown with circles are the reference results from the GENE code, with blue crosses iGENE (same setup as GENE) and in red iGENE with a spectral representation in velocity space. Resolution in velocity space is indicated in the legend.}
    \label{fig:linear_ITG_ae}
\end{figure*}
\begin{figure*}[ht]
    \centering
    \includegraphics[width=0.75\textwidth]{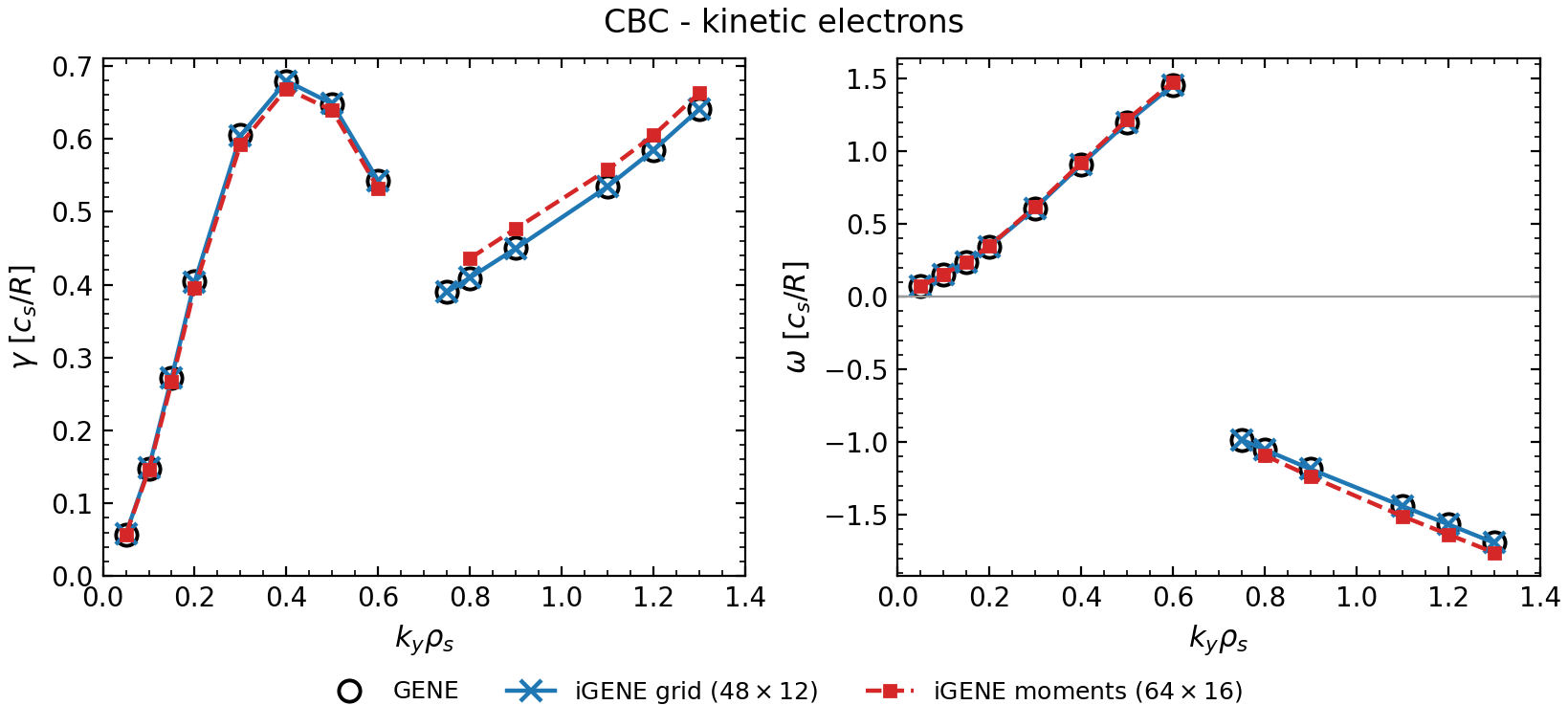}
    \caption{Same as Figure \ref{fig:linear_ITG_ae} but considering kinetic electrons.}
    \label{fig:linear_ITG_ke}
\end{figure*}
We begin by considering the standard cyclone base case (CBC) scenario. We assume circular concentric flux surfaces with inverse aspect ratio $\epsilon=0.18$ and  safety factor $q=1.4$ and magnetic shear $\hat{s}=0.789$. The density (respectively temperature) gradient $R/L_n$=2.22 (respectively $R/L_T$=6.96). Electrons are adiabatic with the same temperature as ions. Simulations use $n_{kx}\times n_{z}=15\times 24$; a velocity space with $L_{v_\parallel} \times L_{\mu}= 3\times 9$ and $n_{v_\parallel}\times n_{\mu}=32\times 8$ is used on the grid, while moments use $N_{H}\times N_{L}=32\times 16$. We refer to Sec.\ref{sec:acceleration_mom} for a definition of the moments-based discretization. Results are shown in Figure \ref{fig:linear_ITG_ae}. The same simulation is then repeated including kinetic electrons (with deuterium mass ratio) and $\beta=10^{-4}$ with results shown in Fig.\ref{fig:linear_ITG_ke}. In both cases close agreement is recovered on both the ITG and TEM branches, confirming the correct implementation of iGENE.
\begin{figure*}[h]
    \centering
    \includegraphics[width=0.75\textwidth]{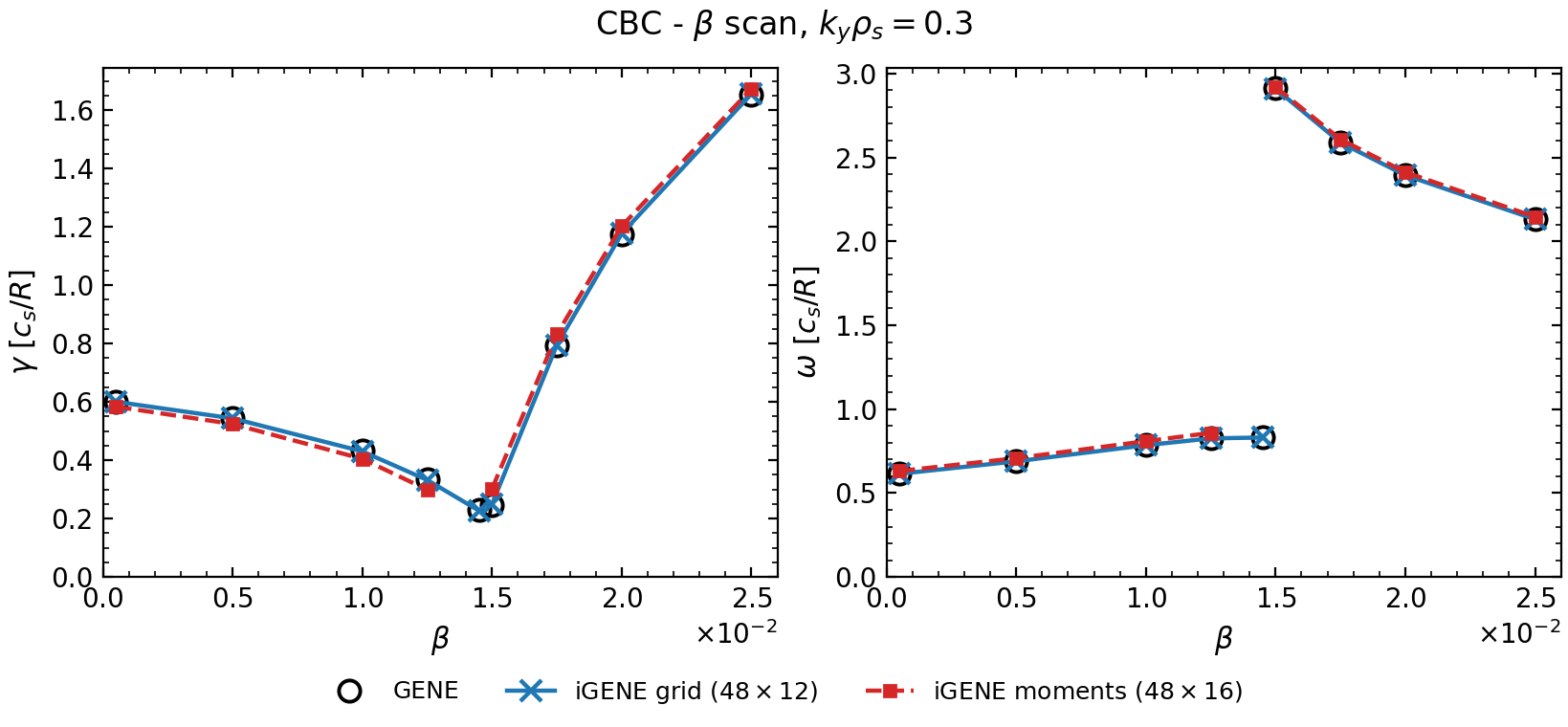}
    \includegraphics[width=0.75\textwidth]{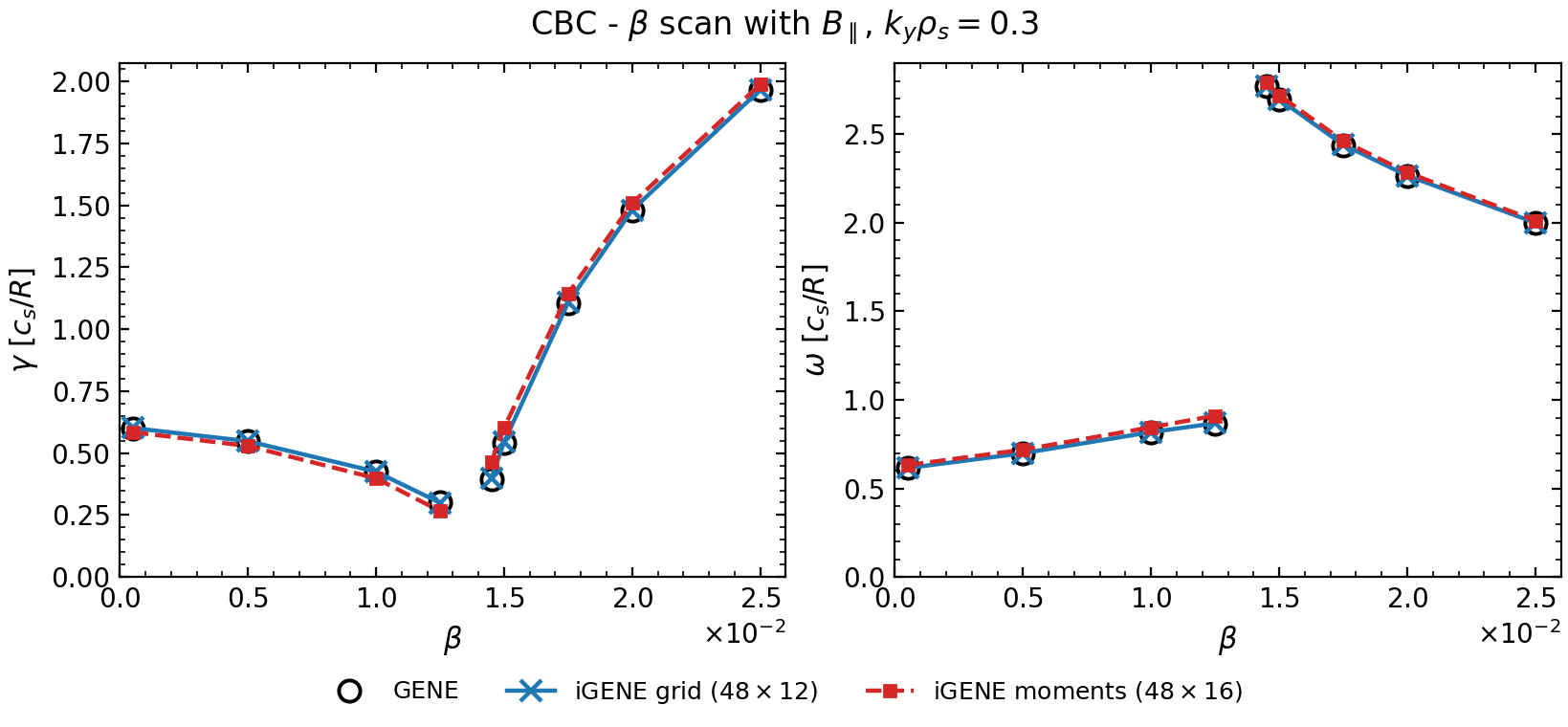}
    \caption{(left) Growth rate $\gamma$ and (right) frequency $\omega$ in units of $c_s/R$ for a $\beta$ scan. Top row results without $B_{\parallel}$ fluctuations, bottom with.  Shown with circles are the reference results from the GENE code, with blue crosses iGENE (same setup as GENE) and in red iGENE with a spectral representation in velocity space. Resolution in velocity space is indicated in the legend.}
    \label{fig:KBM_linear}
\end{figure*}
In order to verify the correctness of the electromagnetic implementation we perform the usual ITG-to-KBM transition as a consequence of increasing $\beta$ \cite{Sheffield_2025, Goerler_2016}. The scan is performed for the mode $k_y\rho_s=0.3$ considering otherwise the same CBC parameters as previously used. Shown in Figure \ref{fig:KBM_linear} is the resulting dispersion relation, with and without including $B_{\parallel}$ fluctuations. Also in this case, all code versions agree, showing a transition from an ITG to a KBM branch (as can be inferred from the jump in frequency) around $\beta=1.5\%$. As expected, including $B_{\parallel}$ lowers the KBM onset and increases its growth rate.\footnote{We remark that the actual values differ from what shown in Figure 2 of Ref.~\cite{Sheffield_2025} because the parameters they use differ from our set, in particular $q$ and $\hat{s}$.}


\subsection{Nonlinear physics}
\begin{figure*}[h!]
    \centering
    \includegraphics[width=0.75\textwidth]{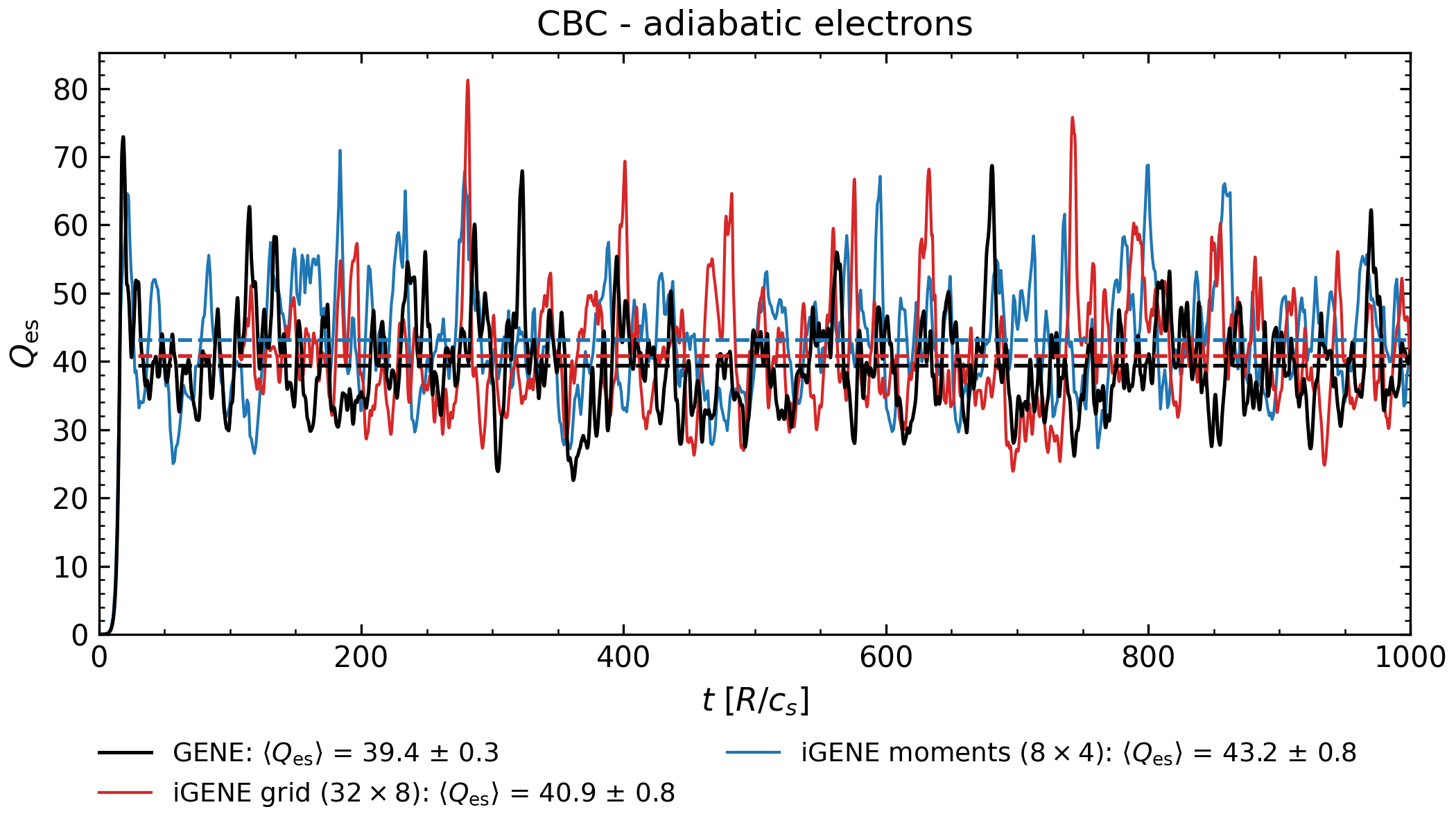}
    \caption{Comparison of the simulated heat flux for CBC parameters with adiabatic electrons. Shown in black are the reference results from the GENE code, in red  iGENE (same setup as GENE) and in blue iGENE with a spectral representation in velocity space. Resolution in velocity space and time averaged values are indicated in the legend.}
    \label{fig:CBC_nonlinear_ae}
\end{figure*}
\begin{figure*}[h!]
    \centering
    \includegraphics[width=0.75\textwidth]{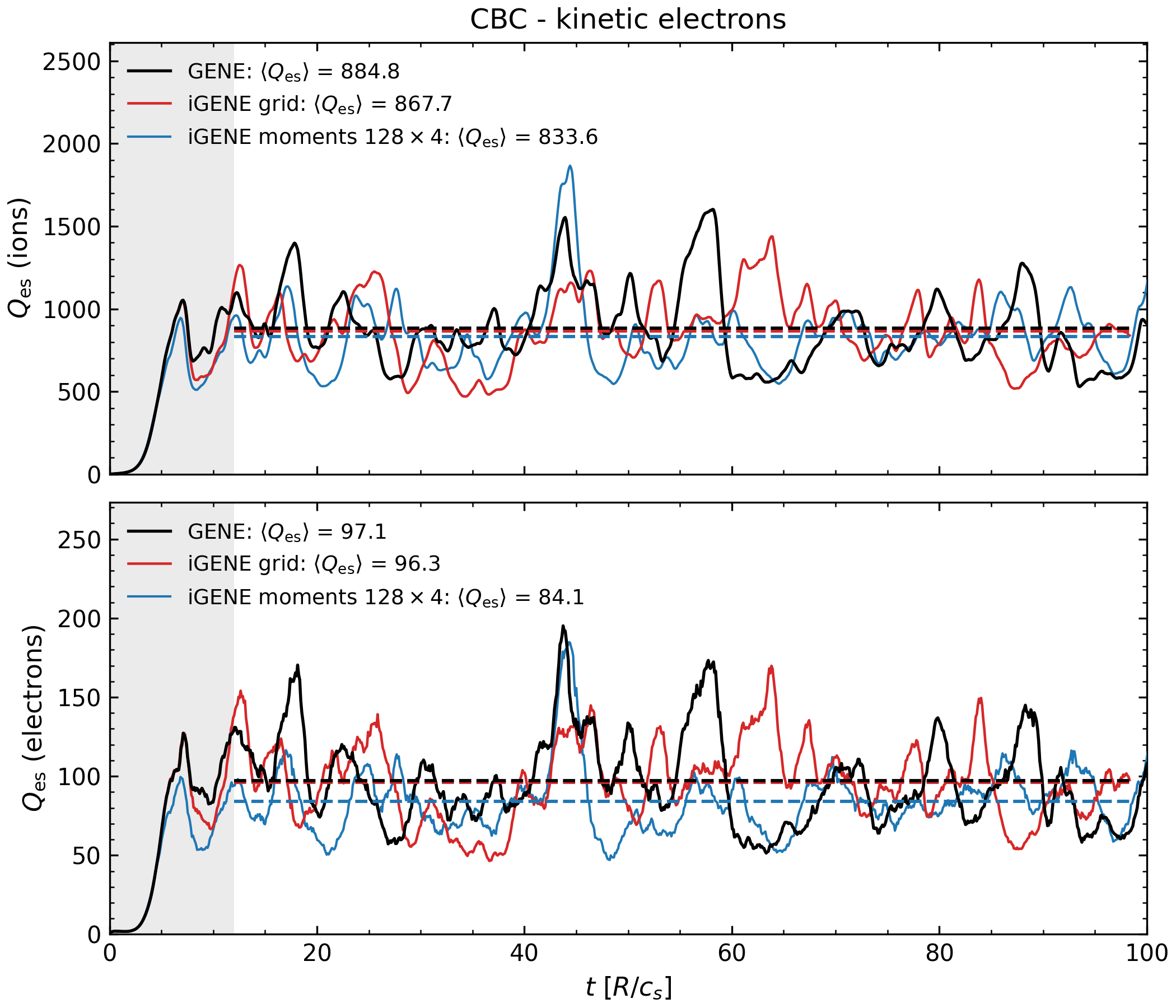}
    \caption{Same as \ref{fig:CBC_nonlinear_ae} but showing results for kinetic electrons simulations.}
    \label{fig:CBC_nonlinear_ke}
\end{figure*}
Nonlinear verification is carried out comparing time averages of the simulated heat fluxes. Error bars on time averages are computed via block averaging: the trace over the saturated window is split into 8 consecutive blocks of equal length, the reported mean is the average of the 8 block means and the error is the standard error of the block means. The results for the CBC parameters with adiabatic electrons are shown in Figure \ref{fig:CBC_nonlinear_ae} whereas kinetic electrons results are depicted in Figure \ref{fig:CBC_nonlinear_ke}. With adiabatic electrons, GENE gives $\langle Q_i\rangle = 39.4\pm0.3$, the iGENE grid version $40.9\pm0.8$ ($+3.8\%$, $1.8\sigma$) and the moment version with $N_H\times N_L=8\times4$ gives $43.2\pm0.8$ ($+9.6\%$, about $4.5\sigma$, with $\sigma$ the combined standard error). The grid version is therefore statistically consistent with GENE, while the low-order moment run carries a small but significant bias, which is reduced at higher moment resolution (Sec.~\ref{sec:acceleration_mom}). With kinetic electrons, the iGENE grid version reproduces the GENE ion and electron heat fluxes within $1.9\%$ and $0.8\%$, respectively. For the moment version a large number of moments, $N_H\times N_L=128\times4$, was used in this case to also converge the electron temperature fluctuations (see the discussion at the end of Sec.~\ref{sec:acceleration_mom}); the ion heat flux is then within $6\%$ of GENE, while the electron heat flux is $13\%$ lower.

\subsection{Gradients of quantities of interest}
After verifying the correctness of the code as an initial value solver, we discuss here its capability to compute gradients of quantities of interest. Given the known challenges with nonlinear gradients due to chaotic divergence of tangents, we limit the comparison to linear scenarios. This verification remains extremely valuable since a dropped dependence, such as a prefactor built from a non-traced value will produce a gradient that is smooth, plausible but wrong. 

We consider the same ITG test case as \cite{Artigues_2026}: $s$--$\alpha$ geometry with $q_0 = 1.4$, $\hat{s} = 0.789$, $\epsilon = r/R_0 = 0.18$, with collisionless adiabatic electrons and CBC gradients. The grid uses $15 \times 32 \times 32 \times 8$ grid in $(k_x, z, v_\parallel, \mu)$ and $k_y \rho_s = 0.1$. 

We evaluate the gradient of the linear growth rate with respect to the same five parameters used in Ref.~\cite{Artigues_2026}: $\{\omega_T, \omega_n, q_0, \hat{s}, \epsilon\}$, with the difference that JAX allows us to evaluate gradients in forward mode. This is in fact the natural choice for the problem at hand. With five inputs and one output it costs five tangent sweeps, but it carries no activation tape, so its memory footprint is independent of the window length $N$ and windows of $10^5$ steps are reachable, contrary to the backpropagation that was originally used in TensorFlow and which was limited by its memory requirements. 
\begin{figure*}[h]
  \centering
  \includegraphics[width=\textwidth]{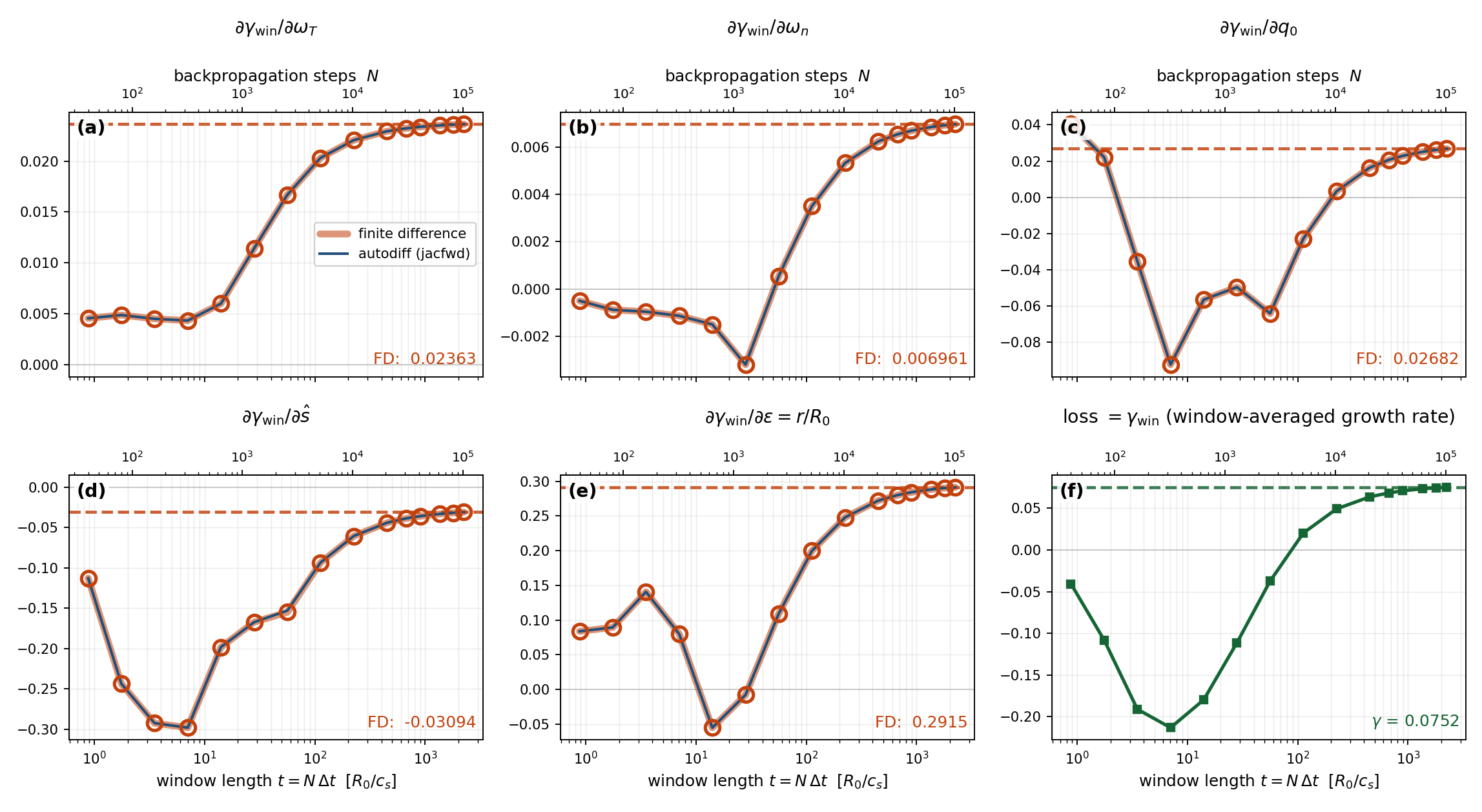}
  \caption{%
    Verification of the differentiated solver on the linear ITG case of
    Ref~\cite{Artigues_2026}. Panels (a)--(e): derivative of the window-averaged growth rate $\gamma_{\mathrm{win}}$ (the growth rate averaged over the window) with respect to each of the five parameters, as a function of the length of the differentiated window. Thick orange line and open circles indicate central finite differences of the growth rate, whereas the thin blue line the  forward-mode autodiff. The dashed line and inset value: indicate the reference value obtained via finite-difference result at the longest window. Panel (f): the actual growth rate. The lower axis is the physical time while the upper axis the corresponding number of forward-propagated steps $N$. }
  \label{fig:grad_convergence}
\end{figure*}
Figure~\ref{fig:grad_convergence} shows the outcome for sixteen window lengths
spanning $N = 40$ to $N = 102\,400$, i.e.\ $t = N\Delta t$ from $0.88$ to
$2247\,R_0/c_s$. Autodiff and finite differences agree at every window and for
every parameter: at the longest window the relative discrepancy is $3.7\times10^{-10}$ ($\omega_T$), $2.8\times10^{-9}$ ($\omega_n$), $1.2\times10^{-7}$ ($q_0$), $1.6\times10^{-4}$ ($\hat{s}$) and $8.3\times10^{-9}$ ($\epsilon$); the largest discrepancy anywhere in the scan is $1.3\times10^{-2}$, on $\hat{s}$, and it decreases monotonically with $N$ beyond $t \simeq 56 R/c_s$.
In particular $\hat{s}$ and $\epsilon$, the two parameters for which Ref.~\cite{Artigues_2026} reported a persistent $30$-$50\%$ offset between autodiff and finite differences, attributed to mesh construction lying outside the differentiable framework, are here reproduced at the same level as the other three. The radial box is indeed now rebuilt from the traced $\hat{s}$ inside the loss, so the mesh dependence is part of the derivative rather than omitted from it.

\section{Code parallelization and performance}\label{sec:performance}
The support of multiple GPUs in iGENE is currently achieved without MPI calls. These were in fact one of the limiting factors of the work in Ref.~\cite{Artigues_2026} where reductions were the only supported collective but were found to be very fragile. In the current implementation, every collective is a JAX collective emitted inside the compiled step. The devices are arranged in a four-dimensional mesh with axes $(b, s, \mu, z)$. $b$ here indicates an additional optional batch dimension, which may or may not be used depending on the simulation setup, to e.g. have a global loss across different flux-tubes. The extent $n_b n_s n_{\mu} n_z$ must equal the device count and each divide the corresponding grid. Axes left unset are resolved automatically, greedily in the order $b \to s \to \mu \to z$. This choice follows our experience with the GENE code on GPU. The auto-tuning capability of GENE, i.e. selecting the best domain decomposition among all possible at run time, is currently not available, mainly because of the repeated compilation cost it would imply. 

The distribution function is laid out as $g[\,b, s, \mu, v_\parallel, z, k_y, k_x]$ and is sharded on $(b, s, \mu, z)$ only; the fields, $[\,b, i\_field, z, k_y, k_x]$, are sharded on $(b, z)$. The parallel velocity and both perpendicular wavenumbers are therefore replicated on every rank by construction. The first choice is made to avoid communicating the corresponding distribution function ghost points when evaluating the parallel derivative in the mirror term. We have to exchange similar large sized ghost cells in the $\mu$ direction when including collisions, but this is done only in the collisional split step, and is thus acceptable. The nonlinearity, which is evaluated pseudo-spectrally and requires a full $k_x$--$k_y$ plane, and the $v_\parallel$ stencil both operate entirely on rank-local data. No transpose and no all-to-all occurs anywhere in the time step. 

The host never drives individual steps. The time loop is a Python loop over blocks, each block a single \texttt{shard\_map(lax.scan(step\_fn))} kernel that advances $K$ steps and emits the per-step diagnostics, with $K$ the diagnostic output interval and the evolved state donated to the kernel. Host-device synchronization thus occurs once per block rather than once per step, and the collectives are emitted inside the compiled region where the compiler can overlap them with computation. Compilation is paid once per block signature and cached.\\

We measure the strong scaling of iGENE on a nonlinear, electromagnetic flux-tube
simulation of the Cyclone Base Case with kinetic electrons, with and without  collisions (Landau operator).

The measurements were made on the Pitagora cluster (CINECA), with four NVIDIA H100 GPUs per node and one process per GPU. For each layout, the code is compiled and then advanced by $150$ and by $650$ steps in the same process; the time per step is $t_\mathrm{step}=[t(650)-t(150)]/500$, which excludes compilation and initialization. The problem does not fit in the memory of a single GPU, so the reference is the two-GPU run with one species per GPU; speed-up and parallel efficiency are defined as
\begin{equation}
S_N=\frac{t_\mathrm{step}(2)}{t_\mathrm{step}(N)},\qquad
E_N=\frac{2\,t_\mathrm{step}(2)}{N\,t_\mathrm{step}(N)} .
\end{equation}

\begin{figure}[ht]
\centering
\includegraphics[width=\textwidth]{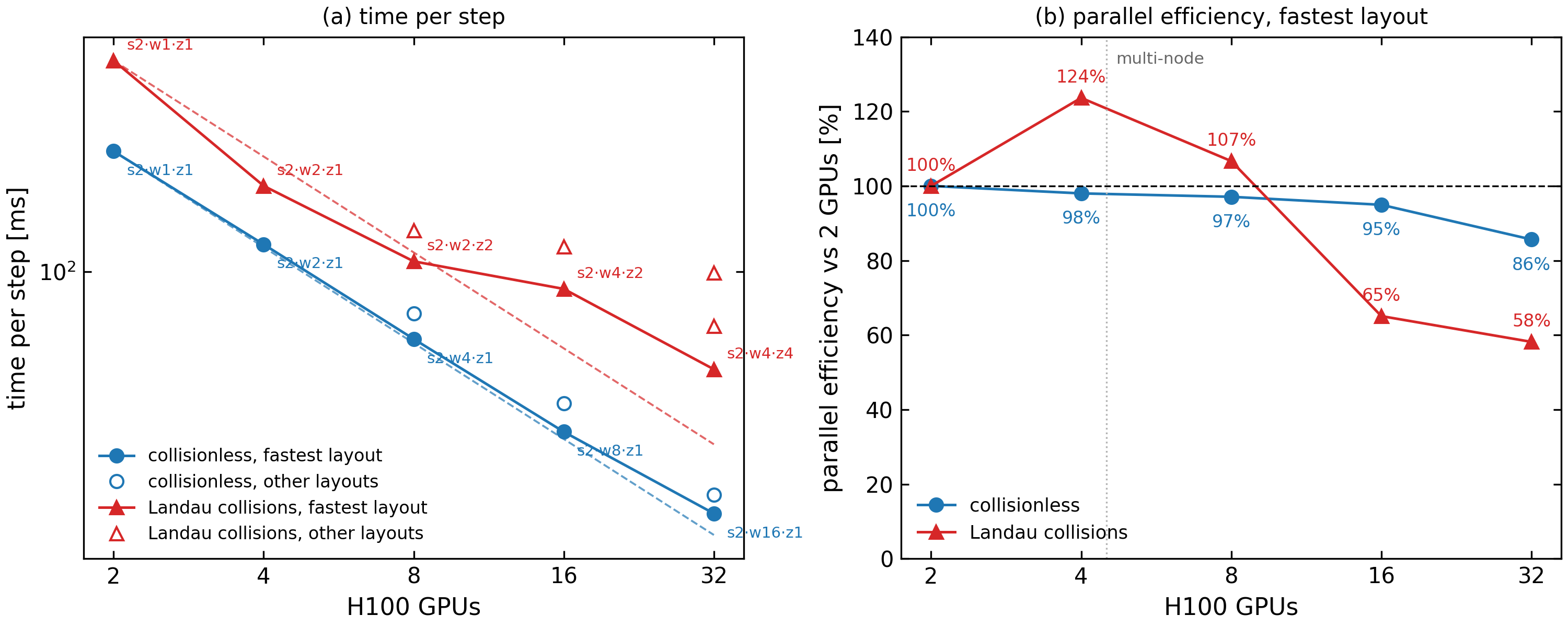}
\caption{Strong scaling of iGENE on H100 GPUs for the electromagnetic Cyclone Base Case with kinetic electrons ($256\times32\times48\times48\times16$), without collisions (circles) and with the Landau collision operator (triangles). (a) Time per step: filled symbols and solid lines indicate the fastest decomposition at each GPU count (labelled $s\cdot w\cdot z$, number of ranks over species, $\mu$ and $z$); open symbols alternative decompositions. Ideal scaling from two GPUs is shown with a dashed line. (b) Parallel efficiency of the fastest decomposition relative to two GPUs; the dotted line separates single-node (up to four GPUs) from multi-node runs.}
\label{fig:scaling}
\end{figure}

Without collisions the code scales almost ideally as seen in Figure~\ref{fig:scaling}. Distributing species and $\mu$, the time per step drops from 238.4\,ms on two GPUs to 17.4\,ms on 32 GPUs over eight nodes, a speed-up of 13.7 (86\% efficiency); up to 16 GPUs the efficiency stays at or above 95\%, so crossing node boundaries has little cost. The decrease at 32 GPUs coincides with a single $\mu$ point per rank. At equal GPU count, replacing part of the $\mu$ split by a $z$ split is 15-23\% slower: the halo exchange of the parallel derivatives is more expensive than the velocity-space reductions. With the Landau operator the time per step is 1.9 times the collisionless one on two GPUs, and the preferred decomposition changes. The collision operator couples all $\mu$ points, so splitting $\mu$ now carries most of the communication cost: at 32 GPUs the decomposition $2\cdot16\cdot1$ takes 98.9\,ms per step, twice as long as $2\cdot4\cdot4$ (49.3\,ms). With the fastest decomposition the speed-up from two to 32 GPUs is 9.3. 



Scaling over a larger number of devices for e.g., multiscale applications\cite{Merlo_2026} may require additional sharding directions. We leave this to future work. 

\section{Algorithmic code acceleration}
\label{sec:acceleration}
In this section we explore novel algorithmic approaches used in iGENE to accelerate the time to solution (besides faster hardware). We will consider both same fidelity approaches as well as options to purposely reduce the fidelity of the results, so that iGENE can be used in multifidelity settings \cite{IPAM}.

\subsection{Multi-rate sub-cycling of the parallel operator}
\label{sec:subcycling}
In an explicit solver the time step is set by the fastest timescale present in the system. With kinetic electrons the binding linear constraint can in certain cases be the parallel electron transit $v_{\mathrm{th},e}\partial_z$ which scales as $(m_i/m_e)^{1/2}$ and tightens with increasing parallel resolution. If this term dominates, the whole right-hand side is advanced at the electron parallel CFL, with the nonlinearity and  collision operator  re-evaluated at the faster rate than they need. A conventional remedy is to treat parallel streaming implicitly, as in \cite{Barnes_2019}. In this case the parallel terms are completely removed from the CFL and the step is set by the perpendicular dynamics alone.
We implemented this approach and it reproduces the explicit results to the splitting order in the electrostatic limit, but for the default electromagnetic cases it did not yield a wall-clock gain on more than one device, mainly because the twist-and-shift chains must be gathered along the full parallel domain at every step; the details are given in Appendix~\ref{App:implicit}. 
We note that the benefit available to any implicit parallel treatment is bounded from the outset: the explicit $E\times B$ nonlinearity retains its own CFL condition, so once the parallel constraint is lifted the step is set by the perpendicular dynamics, which experience shows can be comparable to the fast electron one \cite{Doerk_2014}. As a working alternative, iGENE implements parallel sub-cycling, with a variable number of sub-steps to achieve the maximum possible acceleration at the lowest computational cost.
Parallel sub-cycling is implemented splitting the right-hand side into a \emph{fast} parallel operator $\mathcal{F}$ and a \emph{slow} perpendicular one  $\mathcal{S}$.  $\mathcal{F}$  contains the $\{z,v_\parallel\}$ stencils (streaming and mirror force), the parallel field drive $\partial\chi/\partial z$, and the additional needed field solves. The slow one on the other hand contains curvature and $\nabla B$ drifts, perpendicular field drive, perpendicular hyperdiffusion and the $E\times B$ nonlinearity. Collisions remain split by default using a first-order Runge-Kutta-Chebyshev scheme, so that in collisional runs the composite scheme is formally first order. 

We write the gyrokinetic equation as $\partial_t g = \mathcal{S}(g) + \mathcal{F}(g)$, where both operators include the field solve required to evaluate them, and $\mathcal{S}$ is nonlinear through the $E\times B$ term. Let $\Phi^{\mathcal{X}}_{\tau}$ denote the numerical flow map that advances $\partial_t g=\mathcal{X}(g)$ over a time $\tau$ with one explicit Runge--Kutta step. One outer step of size $\Delta t$ then reads, in the second-order Strang form~\cite{Strang_1968},
\begin{equation}
  g^{n+1} = \Phi^{\mathcal{S}}_{\Delta t/2}\circ
            \bigl(\Phi^{\mathcal{F}}_{\Delta t/N}\bigr)^{N}\circ
            \Phi^{\mathcal{S}}_{\Delta t/2}\,(g^{n}),
  \label{eq:strang}
\end{equation}
i.e.\ a half slow step, $N$ fast sub-steps of size $\Delta t/N$ and a second half slow step, and in the first-order Lie form 
\begin{equation}
  g^{n+1} = \bigl(\Phi^{\mathcal{F}}_{\Delta t/N}\bigr)^{N}\circ
            \Phi^{\mathcal{S}}_{\Delta t}\,(g^{n}),
  \label{eq:lie}
\end{equation}
with local splitting errors $O(\Delta t^3)$ and $O(\Delta t^2)$, respectively, proportional to the commutator of $\mathcal{S}$ and $\mathcal{F}$. Every sub-integration is explicit Runge-Kutta, RK4 by default, whose stability region covers the imaginary axis as the advective streaming requires (an RK3 variant for the fast sub-steps trades one evaluation for a smaller stability radius, can be selected by the user; its effectiveness depends on the balance between sub-cycled step cost and  timestep). The scheme is fully explicit so it reuses the existing distributed collectives unchanged and imposes no new constraint on the parallel decomposition. The major cost stems from the repeated field solves. 

Nonlinearly, iGENE adapts the timestep following the recipe in Ref.~\cite{Doerk_2014}. The timestep is re-estimated each step from a CFL condition containing the instantaneous $E\times B$ velocity, which may grow significantly between the linear phase and saturation. Two modifications are required to the algorithm for sub-cycling. First,
the adaptive estimate is fed the \emph{slow} linear cap $\Delta t^{\mathrm{lin}}_{\mathrm{slow}}$, obtained by removing the parallel streaming, mirror and $\mathrm{hyp}_z$ contributions, so the controller tracks only what $\mathcal{S}$ integrates and the outer step follows the turbulence rather than the electron transit. Second, the fast sub-step must stay resolved. At fixed $N$ the outer step is capped,
\begin{equation}
  \Delta t = \min\bigl(\Delta t^{\mathrm{CFL}}_{\mathrm{slow}}(v_E),\;
                       N\,\Delta t^{\mathrm{lin}}_{\mathrm{fast}}\bigr),
  \label{eq:fixedN}
\end{equation}
guaranteeing $\Delta t/N \le \Delta t^{\mathrm{lin}}_{\mathrm{fast}}$ by construction; alternatively $N$ itself adapts, the outer step taking the full slow CFL with
\begin{equation}
  N = \bigl\lceil \Delta t^{\mathrm{CFL}}_{\mathrm{slow}}(v_E) \big/
                  \Delta t^{\mathrm{lin}}_{\mathrm{fast}} \bigr\rceil .
  \label{eq:adaptN}
\end{equation}
Equation~(\ref{eq:adaptN}) is self-regulating: where the perpendicular CFL is tighter (deep in saturation, large $v_E$) it returns $N=1$ and at that point sub-cycling is not advantageous anymore for the additional field solve; where the parallel constraint dominates (fine $n_z$, realistic mass ratio) $N$ grows to amortize $\mathcal{S}$. Adaptive time stepping interacts with automatic differentiation in two ways. First, the step size depends on the state through the CFL estimate. In iGENE the time step is excluded from differentiation via \texttt{stop\_gradient}, so that derivatives are taken along the trajectory for the realized sequence of time steps. Second, the number of iterations of a loop must be known when the computation is traced for reverse-mode differentiation. A data-dependent number of sub-steps, as in Eq.~\eqref{eq:adaptN}, requires a dynamically terminated loop, which supports forward-mode but not reverse-mode differentiation. Adaptive $N$ is therefore restricted to forward-mode sensitivities and to runs that are not differentiated, while gradients computed in reverse mode use a fixed $N$ with the cap of Eq.~\eqref{eq:fixedN}.
\begin{figure}[h!]
\centering
\includegraphics[width=0.92\textwidth]{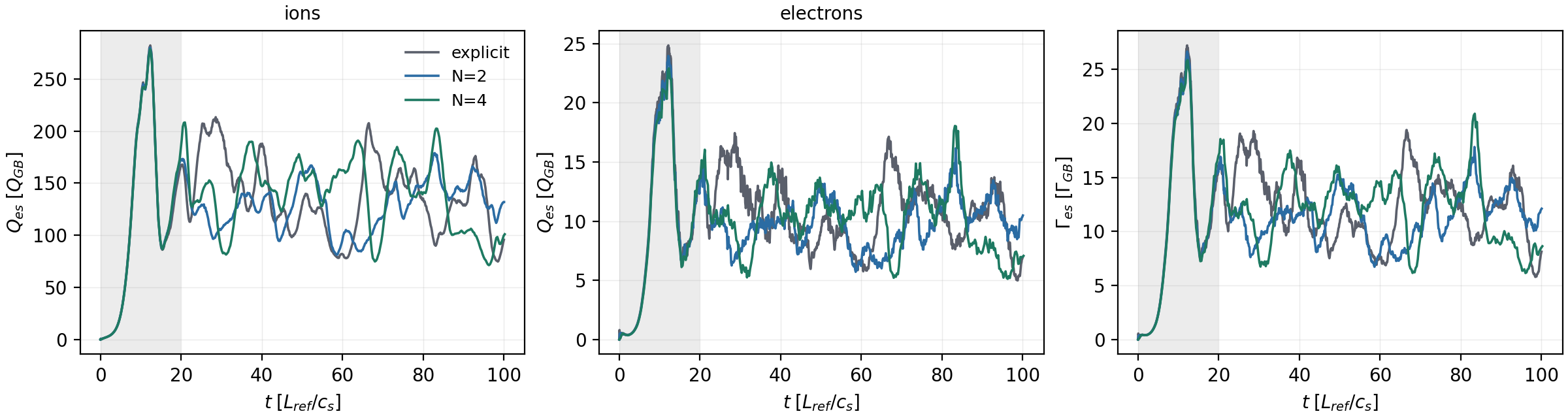}
\caption{From left to right, time traces of ion heat flux, electron heat flux and particle flux. Standard  explicit stepping is shown in black, sub-cycling with N=2 in blue and $N=4$ in green. Statistical agreement is observed for all traces (see also Figure~\ref{fig:subcycle_fluc} in Appendix~\ref{App:convg}).}
\label{fig:subcycle_fluxes}
\end{figure}
An example of the effectiveness of the algorithm is shown in Figures
\ref{fig:subcycle_fluxes} and~\ref{fig:subcycle_gain}, and in
Figure~\ref{fig:subcycle_fluc} in Appendix~\ref{App:convg}. The considered
case is the CBC parameters (circular geometry,  $q_0 = 1.4$, $\hat{s} =
0.789$, $\epsilon = 0.18$, two kinetic species with $m_i/m_e = 3671$,
collisionless and $\beta = 10^{-3}$). To avoid ETG modes we assume a pure ITG
drive, setting $R/L_{T_e} = 0$. The phase-space grid uses  $(n_x, n_{k_y},
n_z, n_{v_\parallel}, n_\mu) = (128, 16, 64, 32, 12)$ over a box  of $L_x =
150.8\,\rho_s$ and $k_{y,\mathrm{min}}\rho_s = 0.1$. The parallel resolution
is purposely kept high, a situation that is often encountered when modeling
strongly shaped tokamak plasmas or
stellarators. Figure~\ref{fig:subcycle_fluxes} compares heat and particle
fluxes for both species, which are found to be statistically
equivalent. Averaging over the quasi steady state region of the run, we obtain
for electrons (respectively ions) without sub-cycling: $Q_{es}$=10.03 $\pm$
0.57 (128.20 $\pm$ 6.95),   9.84 $\pm$ 0.48 (130.15 $\pm$ 5.75) for $N$=2 and
10.25 $\pm$ 0.62 (138.15 $\pm$ 8.07) when $N=4$. Larger values of $N$ are not
possible for the selected case. Beyond $N=4$ the outer step hits the slow
perpendicular $E\times B$) CFL, so Eq.~(\ref{eq:adaptN}) caps $N$.  A similar agreement is found for all relevant fluctuating fields, see Fig.~\ref{fig:subcycle_fluc} in Appendix~\ref{App:convg}. The computational gain, measured as time to solution,  reaches in this case  1.89x for $N=4$, as shown in Figure~\ref{fig:subcycle_gain}.
\begin{figure}[h!]
\centering
\includegraphics[width=0.52\textwidth]{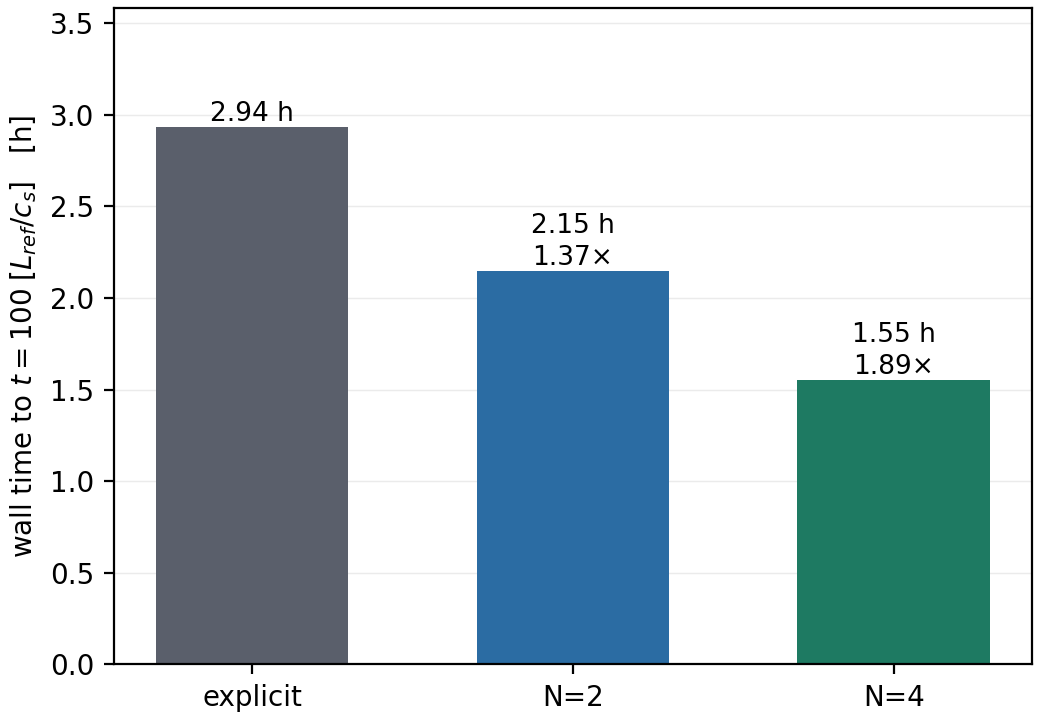}
\caption{Computational gain achieved via sub-cycling the parallel operator for the case described in Figure~\ref{fig:subcycle_fluxes}.}
\label{fig:subcycle_gain}
\end{figure}

\subsection{Mixed precision and tuned GPU FFT kernels}
\label{sec:mixed_precision}

A natural way to accelerate gyrokinetic simulations is to reduce the arithmetic precision. The Fortran version of GENE is commonly run in single precision in production, which roughly halves both the time to solution and the memory footprint \cite{Maurer_2024}. This is possible because the quantities of interest of gyrokinetic simulations are not the details of individual time traces, but their statistical properties, which are considerably more robust. For many applications, the loss of accuracy due to reduced precision is therefore not relevant for the downstream task.

For these reasons, iGENE natively supports both single and double precision. GPUs additionally offer lower-precision formats, typically bfloat16 (bf16) and half precision (fp16), which we explore here to further accelerate the code.\footnote{On the H100, fp8 arithmetic is available only in the tensor cores, i.e.\ for dense matrix products, whereas the dominant iGENE kernels are FFTs, pointwise products and stencils. For these kernels fp8 can therefore serve only as a storage format. Schemes that emulate higher precision using low-precision tensor-core products  do exist~\cite{ozaki_1, ozaki_2}, but they apply to matrix multiplications and would require recasting the transforms as dense matrix products, which we have not explored.} JAX does not support complex types below single precision, so only selected parts of the code can be ported, and a complete port to bf16 is currently not possible. Fortran offers only limited and non-portable support for half-precision types, so similar restrictions apply to the original code \cite{Pfeiler}.

All measurements in this section were performed on a single NVIDIA H100 with 94\,GB HBM3 on the EUROfusion Gateway with JAX 0.10.2 and CUDA 12. The practical HBM bandwidth, measured with a jitted \texttt{y = 2x} over 2\,GB arrays, is 2170\,GB/s for complex64, float32 and bfloat16 (within 1.5\%). Times per step are steady-state values after a warm-up phase, with diagnostics and checkpoints disabled, and are taken either from a dedicated timing pass without recompilation or from the solver's internal block timings. 

\begin{table}[h]
\centering\small
\caption{Simulation cases. All are nonlinear flux-tube runs in circular
geometry ($q_0=1.4$, $\hat s=0.789$, $\epsilon=0.18$) with $R/L_T=6.96$ and
$R/L_n=2.22$ for both species. Resolutions are given as
$n_{k_x}\times n_{k_y}\times n_z\times n_{v_\parallel}\times n_\mu$.}
\label{tab:cases}
\begin{tabular}{@{}lp{0.8\textwidth}@{}}
\hline
case & resolution and physics \\
\hline
A & $256\times32\times16\times48\times12$; 2 species at deuterium mass ratio, $\beta=10^{-3}$, no $E\times B$ shear \\
B & $128\times32\times16\times48\times12$;
    2 species at deuterium mass ratio, $\beta=10^{-3}$,
    $E\times B$ shear rate $10 c_s/R$ \\
C & $256\times32\times24\times48\times12$; 2 species at deuterium mass ratio,
    $\beta=10^{-3}$ \\
\hline
\multicolumn{2}{@{}p{0.95\textwidth}@{}}{}\\
\end{tabular}
\end{table}

Table~\ref{tab:cases} lists the cases used in this section. In case A with fp32 cuFFT transforms, the nonlinear term accounts for 209.7\,ms of a 271\,ms step (77\%). The Poisson bracket is evaluated once per Runge-Kutta stage, and each evaluation requires four inverse two-dimensional transforms and one forward transform of the product. In case B, the bracket accounts for 80\% of the step. The nonlinearity is therefore the first target for optimization, and reducing its precision required replacing the FFT library since vendor libraries do not support the necessary factor of 3 needed for dealiasing.

We developed custom CUDA kernels, called from JAX as XLA foreign-function custom calls, in which the data enter and leave in bf16. The kernels were developed with the assistance of Claude Code. The transform arithmetic runs in fp32 registers in the base kernels, and in packed bf16 in the variant described below. The main design features are:

\begin{itemize}
\item \textbf{Lengths.} $n=32P$ with $P\in\{3,6,12\}$, i.e.\ 96, 192 and 384
along either axis, covering the dealiased grids for
$n_{k_x}\in\{64,128,256\}$ and $n_{k_y}\in\{32,64,128\}$. These are common use case sizes, other values fall back to \texttt{jnp.fft}.

\item \textbf{Warp-resident Cooley--Tukey split.} Each warp transforms one row: a $P$-point DFT in registers (radix 3 for $P=3$, radix $2\times3$ for $P=6$, radix $4\times3$ with a cross-twiddle table for $P=12$), a twiddle multiplication, and a 32-point FFT across the lanes in five radix-2 \texttt{\_\_shfl\_xor\_sync} stages. The two stages whose twiddles are exactly 1 and $-\mathrm{i}$ reduce to a difference and a rotation. Twiddles are precomputed for each $P$ and cached on the device, laid out such that each load is one coalesced 128-byte transaction per warp.

\item \textbf{Real transforms.} The forward real transform packs two real rows into one complex FFT (Hermitian pairing) and stores its output already transposed, so that the transform along the second axis requires no separate transpose. The inverse is the paired complex-to-real transform; the complex inverse uses the conjugation identity.

\item \textbf{Packed bf16 arithmetic.} Real and imaginary parts are stored as two bf16 values in one 32-bit register, so that each butterfly stage requires one shuffle per complex value instead of two, and the arithmetic runs as two-way bf16 SIMD. The compiled code (SASS, \texttt{sm\_90}) of these kernels contains only packed half-precision instructions (\texttt{HADD2},\texttt{HMUL2}, \texttt{HFMA2}) and no scalar floating-point operations. The price is rounding to bf16 after every stage instead of only at input and output.

\item \textbf{Dealiasing pruning.} Only the first $n_{k_y}$ bins along $k_y$ carry data; the padded bins are exact zeros on input and are discarded on output. The kernels skip the corresponding rows, loads and stores. The result is identical to that of the unpruned kernels, up to the sign of exact zeros.
\end{itemize}

The kernels were verified against an fp64 reference transform for all supported lengths and batch shapes. Besides the transforms, the remaining operations in the evaluation of the nonlinearity were also restructured to use reduced precision.

\begin{table}[h]
\centering\small
\caption{Transform- and step-level timings on a single H100. Cases refer to
Table~\ref{tab:cases}.}
\label{tab:fft}
\begin{tabular}{@{}llrrr@{}}
\hline
quantity & case & before (ms) & after (ms) & speed-up \\
\hline
  FFTs: fp32 cuFFT $\to$ bf16 custom              & A &  87.2 &  22.0 & $3.97\times$ \\
  FFTs: fp32 cuFFT $\to$ bf16 custom              & B &  51.6 &  15.1 & $3.42\times$ \\
  Poisson bracket: fp32 cuFFT $\to$ bf16          & A & 209.7 &  62.6 & $3.35\times$ \\
  Poisson bracket: fp32 cuFFT $\to$ bf16          & B & 122.2 &  37.2 & $3.29\times$ \\
  step: fp32 cuFFT $\to$ bf16 bracket             & A & 271.0 & 123.8 & $2.19\times$ \\
  step: fp32 cuFFT $\to$ bf16 bracket             & B & 152.6 &  67.9 & $2.25\times$ \\
  step: fp32 cuFFT $\to$ fp32 custom              & A & 271.0 & 156.2 & $1.74\times$ \\
  step: fp32 cuFFT $\to$ fp32 custom              & B & 152.6 &  74.6 & $2.04\times$ \\
  step: fp32 custom $\to$ bf16 bracket            & A & 156.2 & 123.8 & $1.26\times$ \\
  step: fp32 custom $\to$ bf16 bracket            & B &  74.6 &  67.9 & $1.10\times$ \\
\hline
\end{tabular}
\end{table}

Table~\ref{tab:fft} summarizes the measurements. Comparing the fp32 custom kernels with cuFFT at the same precision isolates the gain due to the implementation: $1.74\times$ and $2.04\times$ per step for cases A and B. Switching the bracket to bf16 adds a further $1.26\times$ and $1.10\times$. Note that part of the implementation gain comes from the dealiasing pruning, i.e.\ from work that a general-purpose library cannot know to skip, rather than from a faster transform algorithm.

The custom transform reaches about 350\,GB/s and 6\,TFLOP/s, i.e.\ about 16\% of the measured memory bandwidth and a small fraction of the arithmetic peak. It is thus limited neither by bandwidth nor by arithmetic throughput, but by the latency of the shuffle chain across the warp. It nevertheless outperforms cuFFT because each row is kept in registers across both factors of the split and the transpose is fused into the store, which removes intermediate passes through HBM. 

The effect on the whole step follows from the fraction of time spent in the bracket. In case A, replacing the bracket alone gives a 3.35$\times$ speedup which reduces to 2.19 over the entire step. In case B, the step improves by $2.25\times$. More generally, we observe that reduced storage precision pays off for data that are written to HBM and read back, such as FFT intermediates, but not for data streams that the compiler already fuses, where the conversions cost more than the saved memory traffic.

\begin{figure}[h!]
\centering
\includegraphics[width=0.5\textwidth]{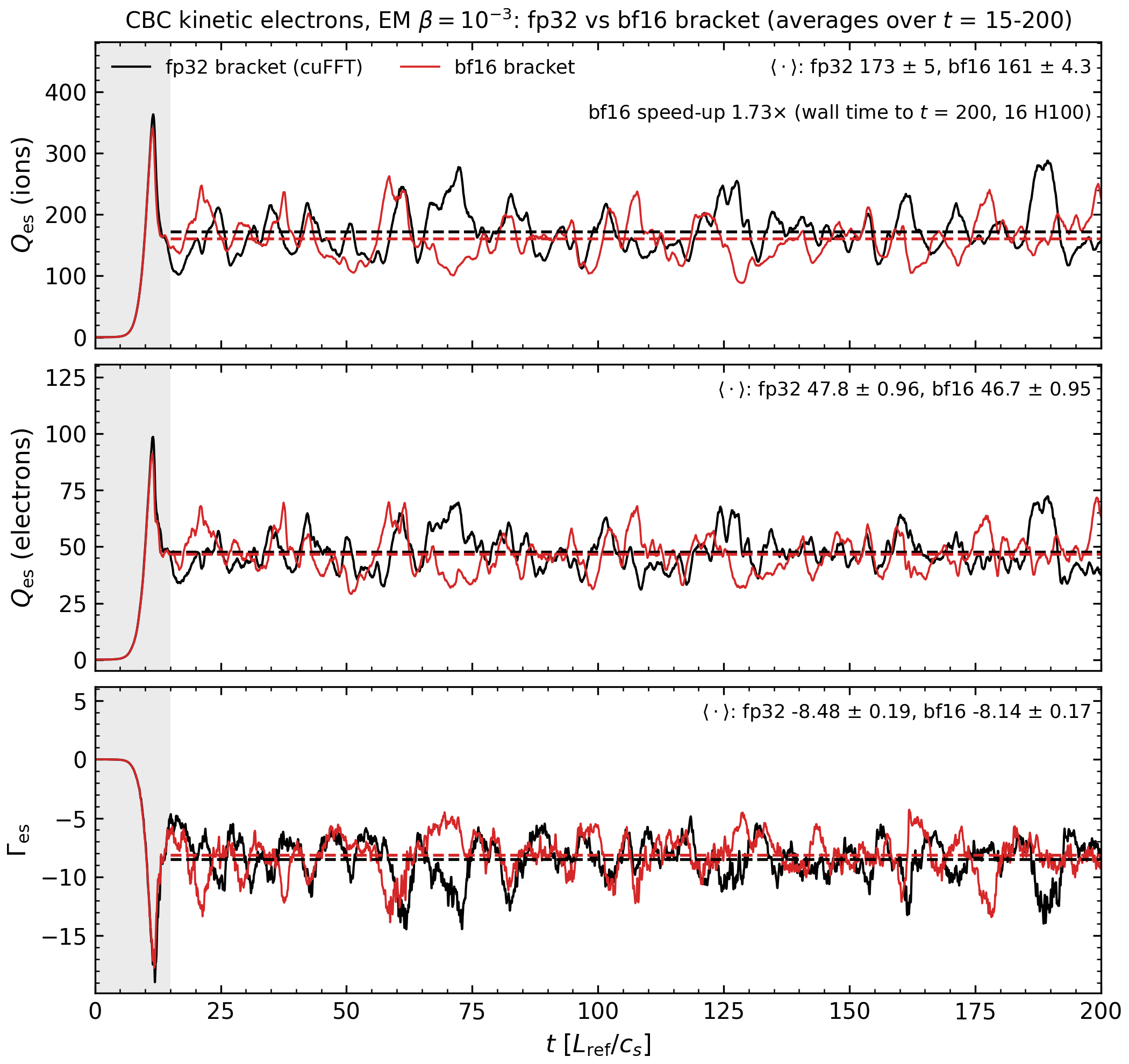}
\caption{Time traces of the simulated heat and particle fluxes for CBC parameters. Black curves indicate results obtained with fp32 arithmetic and red ones with bf16 in the nonlinearity. All averages agree within $2\sigma$, with a measured 1.73$\times$ speedup.}
\label{fig:traces_bfloats}
\end{figure}

Finally, we compare iGENE with GENE on case C, which uses somewhat larger
grids. Results on a single H100 are listed in Table~\ref{tab:gene} as time per
step and the corresponding speedup. GENE was compiled using the default
production settings (release 3.1), and due to memory constraints, a double precision run was not feasible (GENE's estimated memory requirements for double is 105~GB, above the available). 
\begin{table}[h]
\centering\small
\caption{Comparison of GENE and iGENE on case C, single H100.}
\label{tab:gene}
\begin{tabular}{@{}lrrr@{}}
\hline
configuration & s/step & speedup vs GENE & speedup vs iGENE cuFFT \\
\hline
GENE, fp32                         & 0.575 s & 1.00$\times$ & 0.78$\times$ \\
iGENE, fp32, cuFFT                 & 0.447 s & 1.29$\times$ & 1.00$\times$ \\
iGENE, fp32, custom FFT            & 0.267 s & 2.15$\times$ & 1.67$\times$ \\ 
iGENE, bf16 bracket                & 0.219 s & 2.63$\times$ & 2.04$\times$ \\
\hline
\end{tabular}
\end{table}
The speedup of iGENE vs. GENE is due to additional optimizations in iGENE. Some of these can be backported to Fortran, which has not been done here as it requires a significant code refactoring. A comparison of the resulting time traces is shown in Figure \ref{fig:traces_bfloats}. Note that the fluxes are different from what we reported in Figure \ref{fig:CBC_nonlinear_ke} because in this case we are keeping the same temperature and  temperature gradient of ions and electrons.
The time-averaged fluxes agree within $2\sigma$: the ion heat flux is $6.9\%$ lower with bf16 ($173\pm5$ vs.\ $161\pm4.3$), the electron heat flux $2.3\%$ lower ($47.8\pm0.96$ vs.\ $46.7\pm0.95$) and the particle flux $4.0\%$ lower in magnitude ($-8.48\pm0.19$ vs.\ $-8.14\pm0.17$). Since all three shift in the same direction, a small systematic bias of bf16 cannot be excluded from this single run. Error bars are computed via block averaging as described in Sec.~\ref{sec:validation}. Overall, we achieve for this case a 1.73$\times$ speedup, somewhat less than what reported in Table \ref{tab:gene} ($2.04\times$ relative to iGENE with cuFFT, Table~\ref{tab:gene}) because the simulation now also includes communications and the full set of diagnostics.

\subsection{Velocity-space representation}\label{sec:acceleration_mom}
In recent years, representing the distribution function in velocity space by Hermite--Laguerre polynomials has emerged as an effective alternative to grid discretizations, and has been implemented in several codes: GX~\cite{Mandell_2024}, GKX~\cite{Jorge}, GYACOMO~\cite{Hoffmann_2023}, GENE-X~\cite{Frei_2025} and PQLS~\cite{Wilms_axriv_2026}, to name a few. The gyrokinetic equation is replaced by a hierarchy of equations for the Hermite--Laguerre moments, truncated at a chosen order in each index. This makes the representation attractive for multi-fidelity approaches~\cite{IPAM}, since the fidelity of a run is set directly by the number of retained moments. Gyroaverages also become analytic, and the field equations involve only the two lowest Hermite moments, so that the velocity-space integrals reduce to sums over the Laguerre index.

A drawback is the limitation of the time step. A grid in $v_\parallel$  bounds the streaming and drift frequencies. A truncation at $N_H$ Hermite and $N_L$ Laguerre moments instead resolves parallel velocities up to roughly $\sqrt{2N_H}\,v_{\mathrm{th}}$ and perpendicular energies up to roughly $4N_L\,T$~\cite{Mandell_2018}. The largest streaming frequency therefore grows like $\sqrt{N_H}$, and the drift frequencies grow with both indices, so that the explicit time step decreases as the resolution is increased~\cite{Mandell_2024}. Nonetheless, for certain applications the number of moments can be kept small in exchange for a reduced accuracy but potentially  accelerating the time to solution since far fewer degrees of freedom are evolved.

iGENE supports now also a Hermite-Laguerre discretization as a runtime option. Both representations share the entire code infrastructure, including geometry, parallelization, time integration and the evaluation of the $\mathbf{E}\times\mathbf{B}$ nonlinearity.

The derivation of the gyrokinetic equations using a Hermite-Laguerre basis is not new and can be found in several previous works, see e.g. \cite{frei2023moment}. We report it here using iGENE notation for completeness. We define $v = v_\parallel/v_{\mathrm{th}}$, with $v_{\mathrm{th}} = (2T/m)^{1/2}$, and replace the magnetic moment by the normalized perpendicular energy $w = \mu B(z)/T$, so that the background $F_0 = n_0\,\pi^{-3/2}v_{\mathrm{th}}^{-3}\exp(-v^2-w)$ is independent of $z$. Let $p_n = H_n/(2^n n!)^{1/2}$ and $L_l$ be the Hermite and Laguerre polynomials, orthonormal with respect to the weights $\pi^{-1/2}e^{-v^2}$ and $e^{-w}$, respectively. The modified distribution function is expanded as
\begin{equation}
  g = \sum_{n<N_H}\sum_{l<N_L} c_{nl}(z)\, p_n(v)\, L_l(w)\, F_0 ,
  \qquad
  c_{nl} = \frac{1}{n_0}\int\mathrm{d}^3v\; p_n L_l\, g .
  \label{eq:mom_expansion}
\end{equation}

With $b = k_\perp^2 mT/(qB)^2$, the Bessel-function argument is $\lambda = \sqrt{2bw}$, and the Laguerre coefficients of the gyroaverage kernels are analytic:
\begin{equation}
  [J_0]_l = e^{-b/2}\frac{(b/2)^l}{l!} ,
  \qquad
  \Big[\frac{2wJ_1(\lambda)}{\lambda}\Big]_l = [J_0]_l - [J_0]_{l-1} ,
  \qquad
  \sum_{l\ge0}[J_0]_l^2 = \Gamma_0(b) ,
  \label{eq:kernels}
\end{equation}
with $[J_0]_{-1}=0$. The first expression follows from the generating function
$\sum_l e^{-t}t^l L_l(w)/l! = J_0(2\sqrt{tw})$ with $t=b/2$, and the second
from its derivative with respect to $b$. The $B_\parallel$ contribution
therefore requires no special function beyond those already needed for $\phi$.

From Eq.~\eqref{eq:g_to_f}, only the $n=1$ moments differ between $g$ and $f$:
\begin{equation}
  c^{f}_{nl} = c_{nl}
    - \frac{q\,v_{\mathrm{th}}}{\sqrt{2}\,T}\,[J_0]_l\, A_\parallel\,\delta_{n1} .
  \label{eq:g2f}
\end{equation}

We use the field-aligned description of the magnetic field, writing $B=C_{xy}\nabla x \times \nabla y$. $J$ is the Jacobian of the coordinate system. We also introduce the curvature coefficients $K_x=1/C_{xy}\left(\partial B/\partial y- \gamma_2/\gamma_1 \partial B/\partial z \right)$, $K_y=1/C_{xy}\left(\partial B/\partial x- \gamma_3/\gamma_1 \partial B/\partial z \right)$, with $\gamma_1=g^{xx}g^{yy}-g^{xy}g^{xy}$ $\gamma_2=g^{xx}g^{yz}-g^{xy}g^{xz}$ $\gamma_3=g^{xy}g^{yz}-g^{yy}g^{xz}$~\cite{merz2008gyrokinetic}.

In these variables, and with $\{a,b\} = \partial_x a\,\partial_y b - \partial_y a\,\partial_x b$,
$h = g + (q/T)F_0\bar{\chi}$,
\begin{equation}
  \bar{\chi} = J_0\phi - v_\parallel J_0 A_\parallel
             + \frac{T}{q}\frac{w}{B}\,\frac{2J_1(\lambda)}{\lambda}\, B_\parallel ,
  \qquad
  \bar{\chi}_0 = \bar{\chi} + v_\parallel J_0 A_\parallel ,
\end{equation}
and
\begin{equation}
  G = v_{\mathrm{th}}\,\hat{\mathbf{b}}\cdot\nabla z
    = \frac{v_{\mathrm{th}}C_{xy}}{JB} ,
  \qquad
  \mathcal{K}_\alpha = \frac{B}{B^*_\parallel}\frac{T}{qB}\,K_\alpha ,
  \qquad
  \Lambda = \frac{B}{B^*_\parallel}\frac{1}{C_{xy}} ,
  \qquad
  \omega_B = \partial_z \ln B ,
  \label{eq:coeffs}
\end{equation}
the gyrokinetic equation of Section~\ref{sec:implementation}, written in terms of $g$, reads
\begin{equation}
  \begin{aligned}
    \partial_t g =\;
      &- G\, v\, \partial_z f\big|_\mu
       \;+\; \tfrac{1}{2} G\, \omega_B\, w\, \partial_v f
       \;-\; \frac{q}{T}\, G\, v\, F_0\, \partial_z \bar{\chi}_0\big|_\mu \\
      &- \big(w + 2v^2\big)\big(\mathcal{K}_x \partial_x
          + \mathcal{K}_y \partial_y\big) h
       \;-\; \Lambda \big[\omega_n + \omega_T\big(v^2 + w
          - \tfrac{3}{2}\big)\big] F_0\, \partial_y \bar{\chi}
       \;-\; \Lambda \big\{\bar{\chi}, g\big\} \\
      & \;+\; C[f] ,
  \end{aligned}
  \label{eq:gk_full}
\end{equation}
The first line contains parallel streaming, the mirror force and the parallel field drive, i.e.\ the fast operator $\mathcal{F}$ of Sec.~\ref{sec:subcycling}; the second line contains the magnetic (curvature and $\nabla B$) drifts, the drive by the background density and temperature gradients, and the $E\times B$ nonlinearity; the last term is the collision operator. Hyperdiffusion and $E\times B$ shear flows are omitted for brevity. Since $w$ depends on $z$ at fixed $\mu$, with $\partial_z w|_\mu = w\,\omega_B$,
\begin{equation}
  \partial_z\big|_\mu = \partial_z\big|_w + \omega_B\, w\,\partial_w ,
  \qquad
  \frac{\mu}{m}(\partial_z B)\,\partial_{v_\parallel}
  = \tfrac{1}{2} v_{\mathrm{th}}\,\omega_B\, w\,\partial_v ,
  \label{eq:chvar}
\end{equation}
the second identity being the mirror term in Eq.~\eqref{eq:gk_full}.

Projecting Eq.~\eqref{eq:gk_full} onto the basis requires the action of $v$, $v^2$, $w$, $\partial_v$ and $w\partial_w$ on the basis functions. Each basis function couples only to its nearest neighbors:
\begin{equation}
  \begin{aligned}
    v\,p_m &= \sqrt{\tfrac{m+1}{2}}\,p_{m+1} + \sqrt{\tfrac{m}{2}}\,p_{m-1}
      &&\equiv \textstyle\sum_n V_{nm}\, p_n , \\[2pt]
    w\,L_k &= -(k+1)L_{k+1} + (2k+1)L_k - k L_{k-1}
      &&\equiv \textstyle\sum_l E_{lk}\, L_l , \\[2pt]
    \partial_v\big(e^{-v^2} p_m\big) &= -\sqrt{2(m+1)}\,e^{-v^2}\, p_{m+1}
      &&\equiv e^{-v^2} \textstyle\sum_n R_{nm}\, p_n , \\[2pt]
    w\,\partial_w\big(e^{-w}L_k\big) &= (k+1)\,e^{-w}\big(L_{k+1}-L_k\big)
      &&\equiv e^{-w}\textstyle\sum_l W_{lk}\, L_l ,
  \end{aligned}
  \label{eq:mom_matrices}
\end{equation}
together with the pentadiagonal matrix $V^{(2)}$ representing $v^2$. All linear collisionless velocity-space operators are therefore banded.
Only the $n=0$ and $n=1$ moments enter the field equations. Neglecting $B_\parallel$ for clarity, quasineutrality~\eqref{eq:qn} and Amp\`ere's law~\eqref{eq:ampere} become
\begin{align}
  \sum_s q_s n_{0s}\sum_{l<N_L}[J_0]_l\, c_{0l,s}
    &= \sum_s \frac{q_s^2 n_{0s}}{T_{0s}}
       \Big(1-\sum_{l<N_L}[J_0]_l^2\Big)\phi ,
  \label{eq:qn_mom} \\
  \Big[k_\perp^2 + \mu_0\sum_s\frac{q_s^2 n_{0s}}{m_s}
       \sum_{l<N_L}[J_0]_l^2\Big] A_\parallel
    &= \mu_0\sum_s q_s n_{0s}\frac{v_{\mathrm{th},s}}{\sqrt2}
       \sum_{l<N_L}[J_0]_l\, c_{1l,s} .
  \label{eq:ampere_mom}
\end{align}
With $B_\parallel$, the coupled system of Section~\ref{sec:implementation} is obtained in the same way, using the second kernel in Eq.~\eqref{eq:kernels}. The polarization and skin terms are truncated consistently with the moment sums on the other side of the equations, i.e.\ $\sum_{l<N_L}[J_0]_l^2$ replaces the analytic $\Gamma_0(b)$. A gyroaveraged Boltzmann response then produces no spurious charge density at any $N_L$. The exact finite-Larmor-radius response is recovered as $N_L\to\infty$. The grid and moment representations therefore solve formally different equations at finite resolution, which converge as $N_L$ grows. The $\mathbf{E}\times\mathbf{B}$ nonlinearity reuses the pseudo-spectral evaluation and dealiasing in $(x,y)$ of the grid scheme.

Truncation at $(N_H, N_L)$ closes the hierarchy by discarding the couplings to moments outside the retained set. Free energy that would otherwise accumulate at the highest retained indices is removed by the diagonal hypercollision term \begin{equation}
  -\nu_h\Big[\Big(\frac{n}{N_H-1}\Big)^{4}
           + \Big(\frac{l}{N_L-1}\Big)^{4}\Big] c_{nl} ,
  \label{eq:hypercoll}
\end{equation}
which replaces the $v_\parallel$ hyperdiffusion of the grid scheme. The
closures of GX~\cite{Mandell_2024} and GYACOMO~\cite{Hoffmann_2023} are also
available to the user. Two collision operators are implemented in closed form in the basis: the conservative Dougherty operator in the form of Mandell \textit{et~al.}~\cite{Mandell_2024}, and the drift-kinetic, like-species limit of the Sugama operator following Frei \textit{et~al.}~\cite{Frei_2021}.

To quantify the achievable speed-up, we first consider the nonlinear CBC case with
adiabatic electrons of Section~\ref{sec:validation}.
\begin{figure}[h!]
\centering
\includegraphics[width=0.9\textwidth]{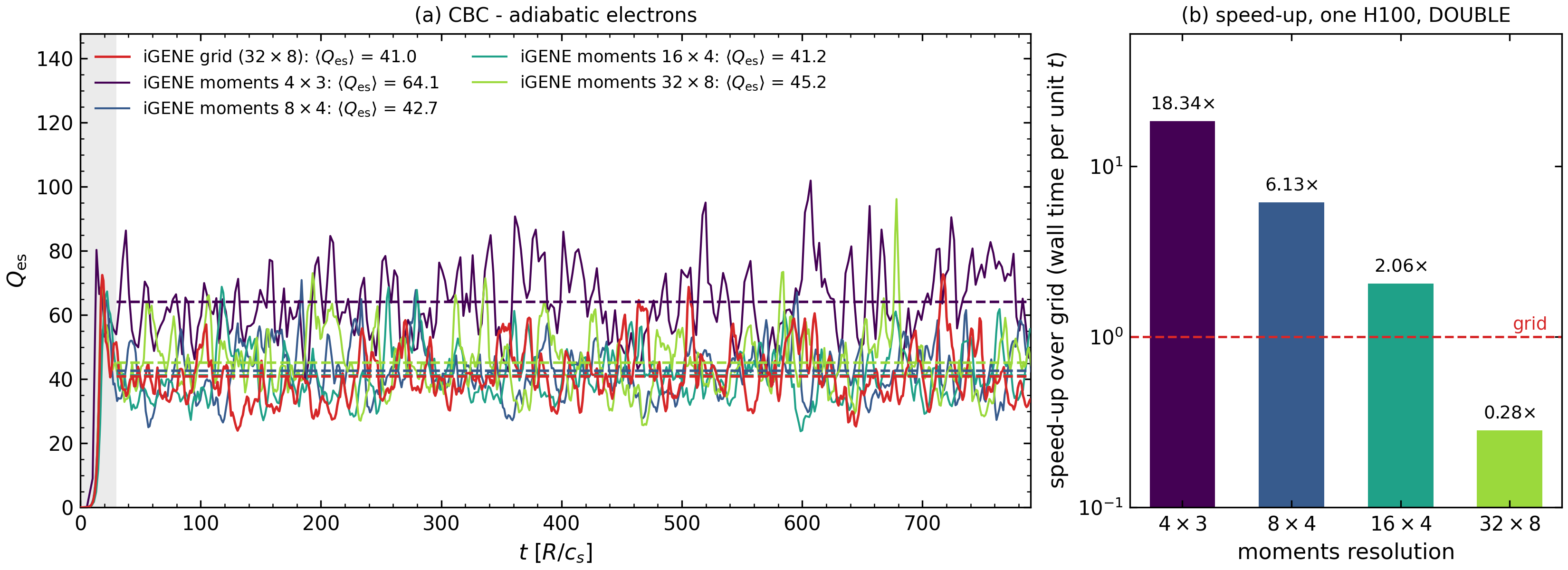}
\caption{Nonlinear CBC with adiabatic electrons. (left) Time traces of the
ion heat flux for different moment resolutions $N_H\times N_L$ and for the
grid reference with $n_{v_\parallel}\times n_\mu = 32\times8$; time-averaged
values are given in the legend. (right) Measured speed-up in time to solution
with respect to the grid reference.}
\label{fig:moments_ae_fluxes}
\end{figure}

\begin{figure}[h!]
\centering
\includegraphics[width=0.9\textwidth]{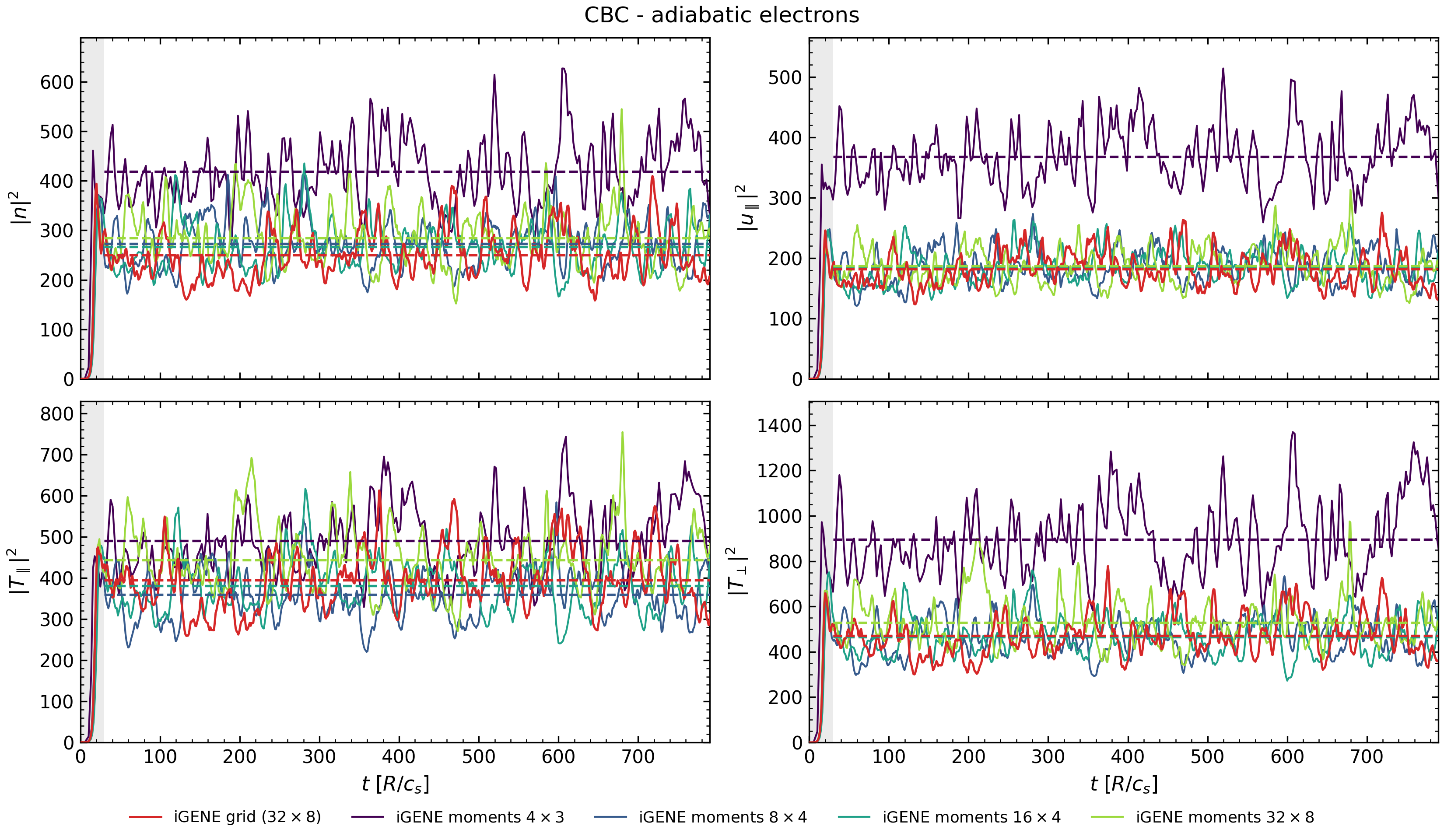}
\caption{Same as Fig.~\ref{fig:moments_ae_fluxes}, but showing temperature, density and parallel velocity fluctuations.}
\label{fig:moments_ae_flucts}
\end{figure}
Figures~\ref{fig:moments_ae_fluxes} and~\ref{fig:moments_ae_flucts} show the
ion heat flux and mean square values of temperature, density and parallel velocity fluctuation for different moment resolutions, compared with the grid reference using $n_{v_\parallel}\times n_\mu = 32\times8$. With $N_H\times N_L = 8\times4$, the time-averaged heat flux, $\langle Q_i\rangle = 42.7$, agrees with the grid result, $\langle Q_i\rangle = 41.0$, while the time to solution is reduced by a factor of $6.1$. This essentially results from a correspondingly lower cost per step, with $32$ instead of $256$ velocity-space degrees of freedom, and an essentially similar time step. We point out that we have not tuned the mesh size in the grid run. We expect that the gap between the two discretizations can be reduced if the box extent and the number of points is optimized~\cite{Goerler_private}. However, such exercise is usually not possible for routine production runs, where ``standard'' resolutions are usually employed. Conversely, the hypercollision strength of the moment runs was not tuned either, and the accuracy of low-order truncations may depend on it. Ultimately the actual speedup will depend on the tolerable accuracy on the downstream task. The speedup must, however, be read together with the accuracy: the $4\times3$ truncation is $18.3\times$ faster than the grid, but gives $\langle Q_i\rangle = 64.1$, i.e.\ a 56\% overestimate. Conversely, the moment representation becomes more expensive than the grid at high resolution: with $N_H\times N_L = 32\times8$, i.e.\ the same number of degrees of freedom as the grid, the time to solution is about four times longer, because of the smaller time step.

The required resolution depends on the quantity of interest. Saturated fluxes converge at a lower number of moments than the linear growth rates of Section~\ref{sec:validation}, which were computed with $32\times16$ moments to resolve all $k_y$, including those near marginal stability. Adiabatic-electron ITG turbulence is particularly favorable for low-order truncations, owing to its smooth velocity-space structure. 
\begin{figure}[h!]
\centering
\includegraphics[width=0.9\textwidth]{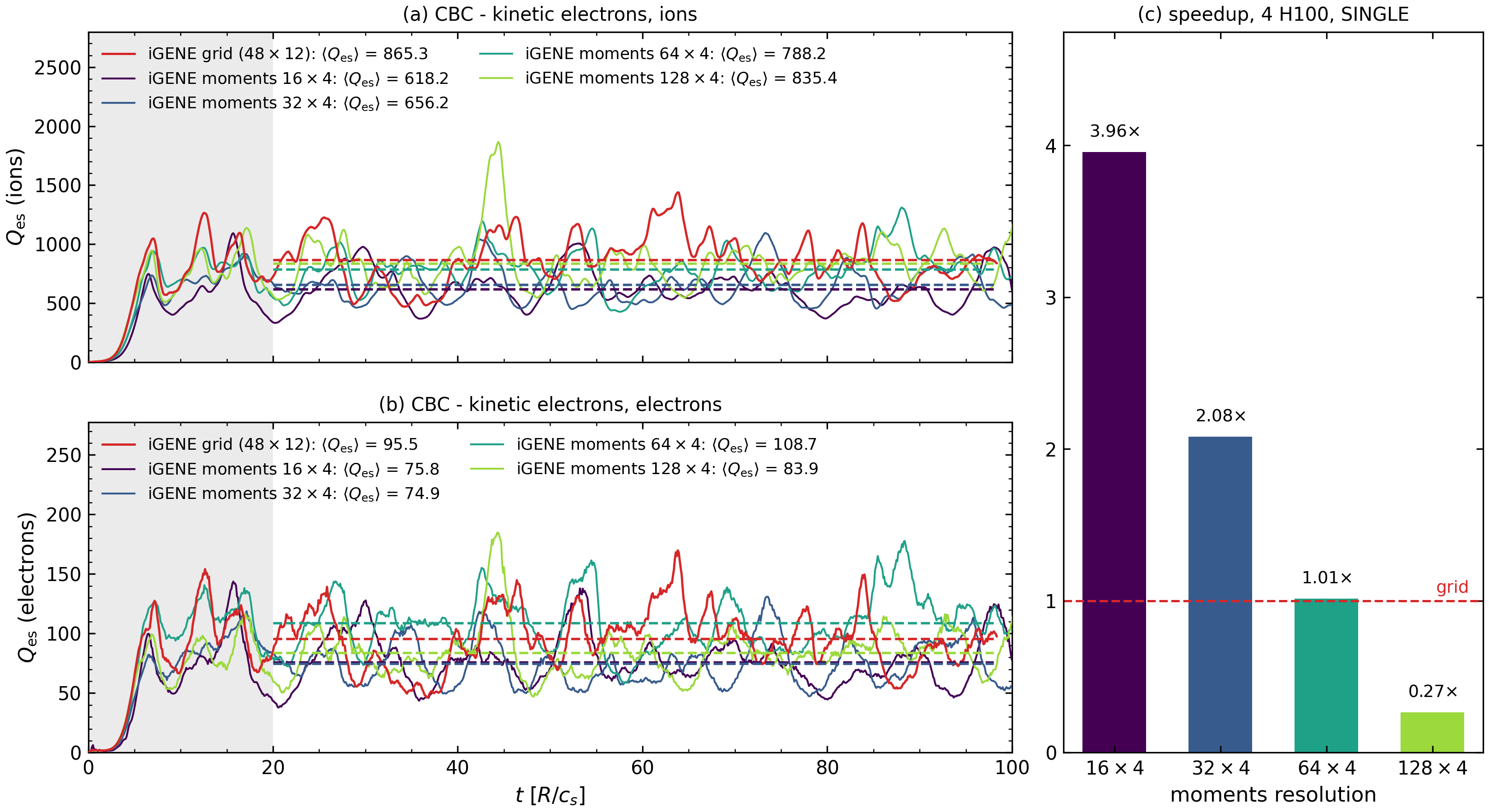}
\caption{Nonlinear CBC with kinetic electrons. (left) Time traces of the
ion heat flux for different moment resolutions $N_H\times N_L$ and for the
grid reference with $n_{v_\parallel}\times n_\mu = 48\times12$; time-averaged
values are given in the legend. (right) Measured speed-up in time to solution
with respect to the grid reference.}
\label{fig:moments_ke_fluxes}
\end{figure}

\begin{figure}[h!]
\centering
\includegraphics[width=0.9\textwidth]{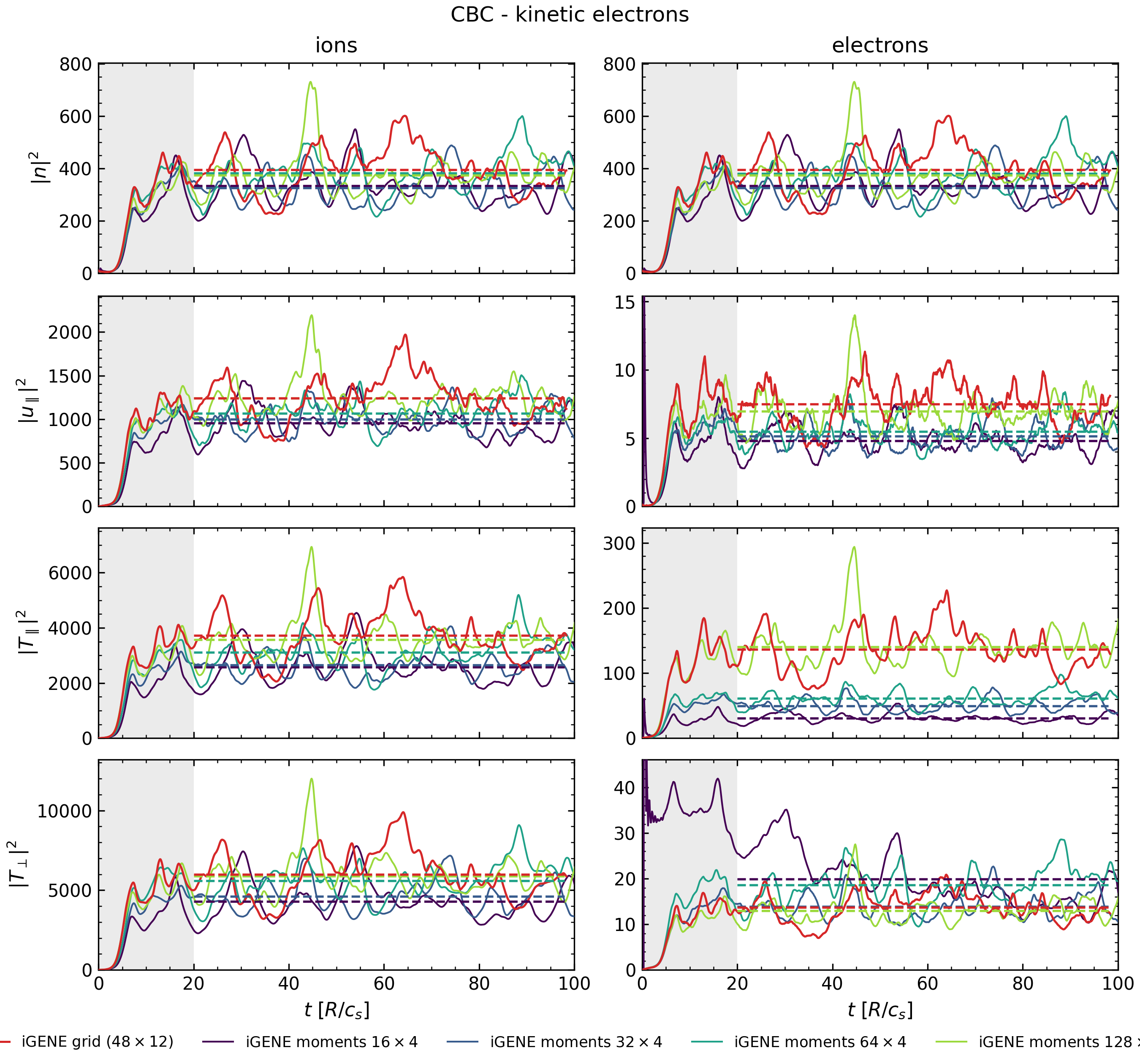}
\caption{Same as Fig.~\ref{fig:moments_ae_fluxes}, but showing temperature, density and parallel velocity fluctuations for ions and electrons. While the heat fluxes converge quickly with the number of moments, temperature fluctuations require a significantly higher resolution, similar to the one needed in linear simulations.}
\label{fig:moments_ke_flucts}
\end{figure}
For this reason, we repeated the same analysis for CBC parameters with kinetic electrons, obtaining the results in Figures \ref{fig:moments_ke_fluxes} and~\ref{fig:moments_ke_flucts} (simulations are in this case done in single precision over 4 H100, to reduce the overall cost). Results are qualitatively similar: at $N_H\times N_L = 16\times4 $ the moment representation is roughly four times faster than the grid version, but the ion flux is under-predicted by about 25$\%$ and the electron temperature fluctuation  under-predicted by more than a factor of two. Statistical match between representation is achieved only at $N_H\times N_L = 128\times4$, which however ends up being roughly four times slower because of the much smaller timestep.

\section {Summary and outlook}\label{sec:conclusion}
We have presented and verified iGENE in its latest version. The code has been fully ported from its original TensorFlow implementation to JAX, keeping differentiability, performance and ease of use as key performance metrics. The code currently features the same local model as its predecessor GENE, namely fully electromagnetic multi-species, collisional flux tube gyrokinetic model. Even though the code runs on multiple GPUs and can immediately benefit from any further speedup from upstream JAX developments, reducing the time to solution remains an essential aspect that needs to be addressed for a turbulence code.
We have therefore investigated several possible speed-up techniques and their impact:
\begin{itemize}
    \item {\it Sub-cycling the parallel dynamics}: owing to the large mass ratio difference between ions and electrons, in many cases parallel dynamics are the limiting factor in linear simulations, a situation worsened for stellarator geometries where the parallel resolution can reach hundreds of points per turn. Nonlinear dynamics can regulate and often further reduce the timestep, however we find that a moderate sub-cycling ratio $N$ can lead to non-negligible code speedups. For the cases we have considered one cannot exceed $N=4$, still accelerating the code by a factor 1.89.
    \item {\it Mixed precision}: reducing precision arithmetic is an obvious way to speed up any calculation. For turbulence simulations this is possible because one is interested in the statistical averages and not exact details of the evolution of the system. Modern GPUs do support reduced precision and are therefore perfectly suited for this goal. However,  the software support is not mature yet so that specific functionalities had to be introduced. JAX itself currently supports only real bfloats, thus reduced precision has been implemented and tested only in the nonlinear term, which is the largest kernel in our runs. With the help of AI, we have developed custom FFT kernels that can be run in bf16 and allow for almost a 4$\times$ speedup compared to vendor libraries in fp32. Overall, this translates to roughly an acceleration by a factor of two of the whole code.    
    \item {\it Velocity space representation}: using a Hermite-Laguerre basis in velocity space has emerged in the last years as a valid alternative to a pure grid based description. To verify its effect and make it available to the user, we have implemented it in iGENE as an alternative runtime option. We find that such a description can significantly outperform the grid based code performance. However this comes at an accuracy cost. In order to produce same accuracy results over an extended range of quantities of interest including fluxes and fluctuations, a large number of moments needs to be retained. This can result in a much smaller timestep thus overall higher computational cost than a grid based description.
\end{itemize}
Overall, these techniques are promising and can lead to an important reduction of the time to solution of a given simulation, hence allow for a faster  answer or more extensive parameter exploration at fixed computational budget.  Certain approaches can be combined, but the gains do not simply multiply, and some techniques act against each other. This is the case for sub-cycling and reduced precision: bf16 accelerates only the $E\times B$ bracket, which belongs to the slow operator $\mathcal{S}$, whereas sub-cycling reduces how often $\mathcal{S}$ is evaluated per unit simulated time. Once sub-cycling is active, a smaller fraction of the run time is spent in the bracket and the bf16 gain shrinks accordingly. Ultimately, the speedup of a simulation is tied to the accuracy of the obtained results, and for many applications a lower one is possible in conjunction with multifidelity approaches. iGENE is now suited for these applications. Next steps include the extension to nonlocal (global) simulations, additional sharding directions for multiscale simulations, backporting the custom FFT kernels and related optimizations to the Fortran GENE code, and the use of the end-to-end gradients in optimization and machine-learning workflows, which will be reported in future publications.

\section{Acknowledgments}
Part of this research was performed while G.M., V.A. and F.J. were visiting the Institute for Pure and Applied Mathematics (IPAM), which is supported by the National Science Foundation (Grant No. DMS-1925919/2422832). G. M. gratefully acknowledges stimulating discussions in the Accelerating Forward Simulations working group at IPAM, in particular M. Golden, G. Holt and T.H Watanabe. C.P.'s work was funded by the German Federal Ministry of Research, Technology and Space within the DaREXA-F project (Grant No. 16ME0644). Simulations were performed on the Pitagora and EFGW clusters at Cineca and on LUMI at CSC.

\section*{Use of AI tools}
Claude Code (Anthropic) was used to assist the development of parts of iGENE, in particular the custom CUDA FFT kernels described in Sec.~\ref{sec:mixed_precision}. All code was reviewed and verified by the authors as described in the text.

\section*{Data availability}
The data supporting the figures (input files, time traces, timing logs) is available from the corresponding author upon reasonable request.

\appendix
\section{Impact of sub-cycling rate $N$ on fluctuations}\label{App:convg}
\begin{figure}[h!]
\centering
\includegraphics[width=0.92\textwidth]{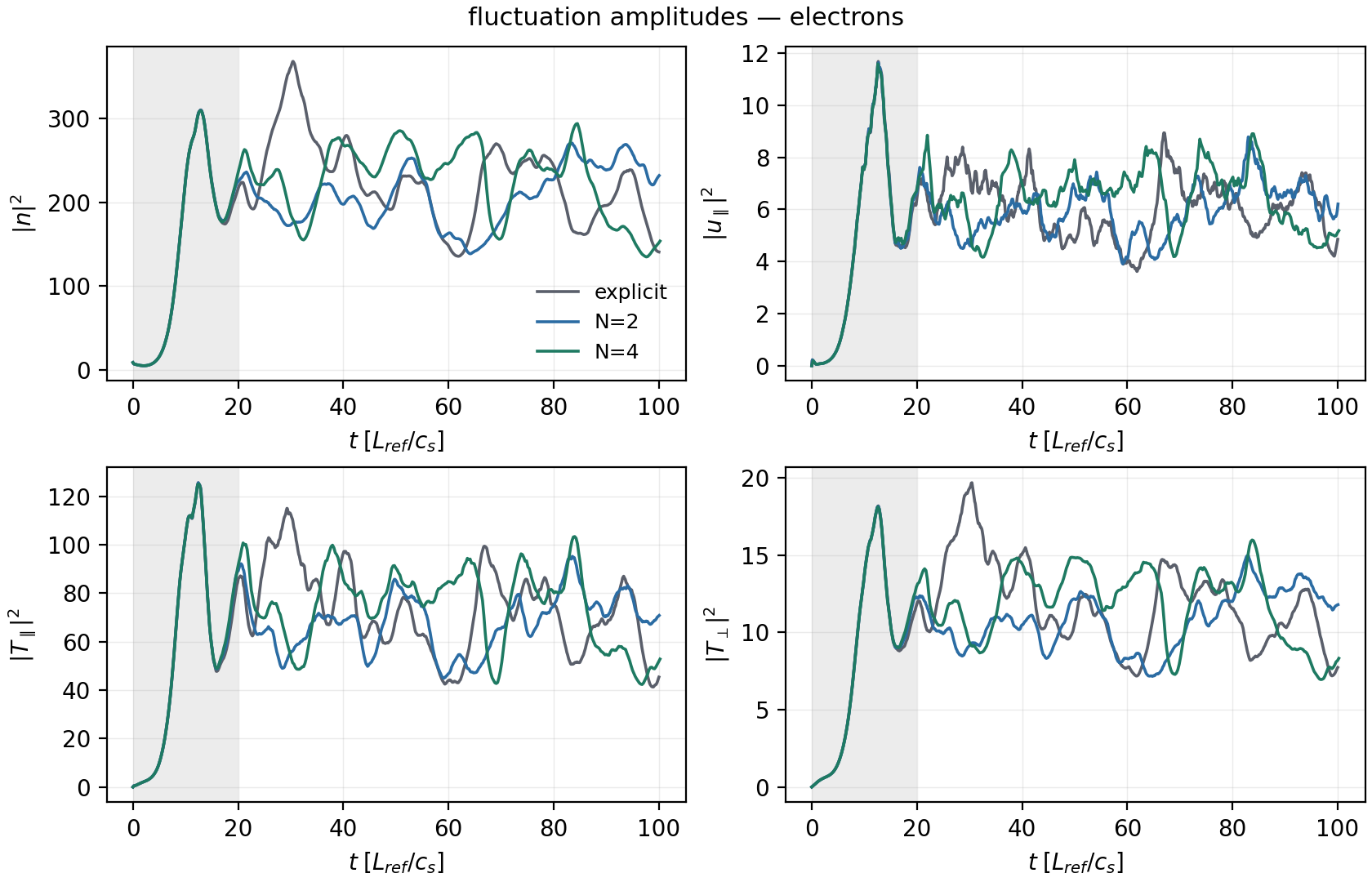}
\includegraphics[width=0.92\textwidth]{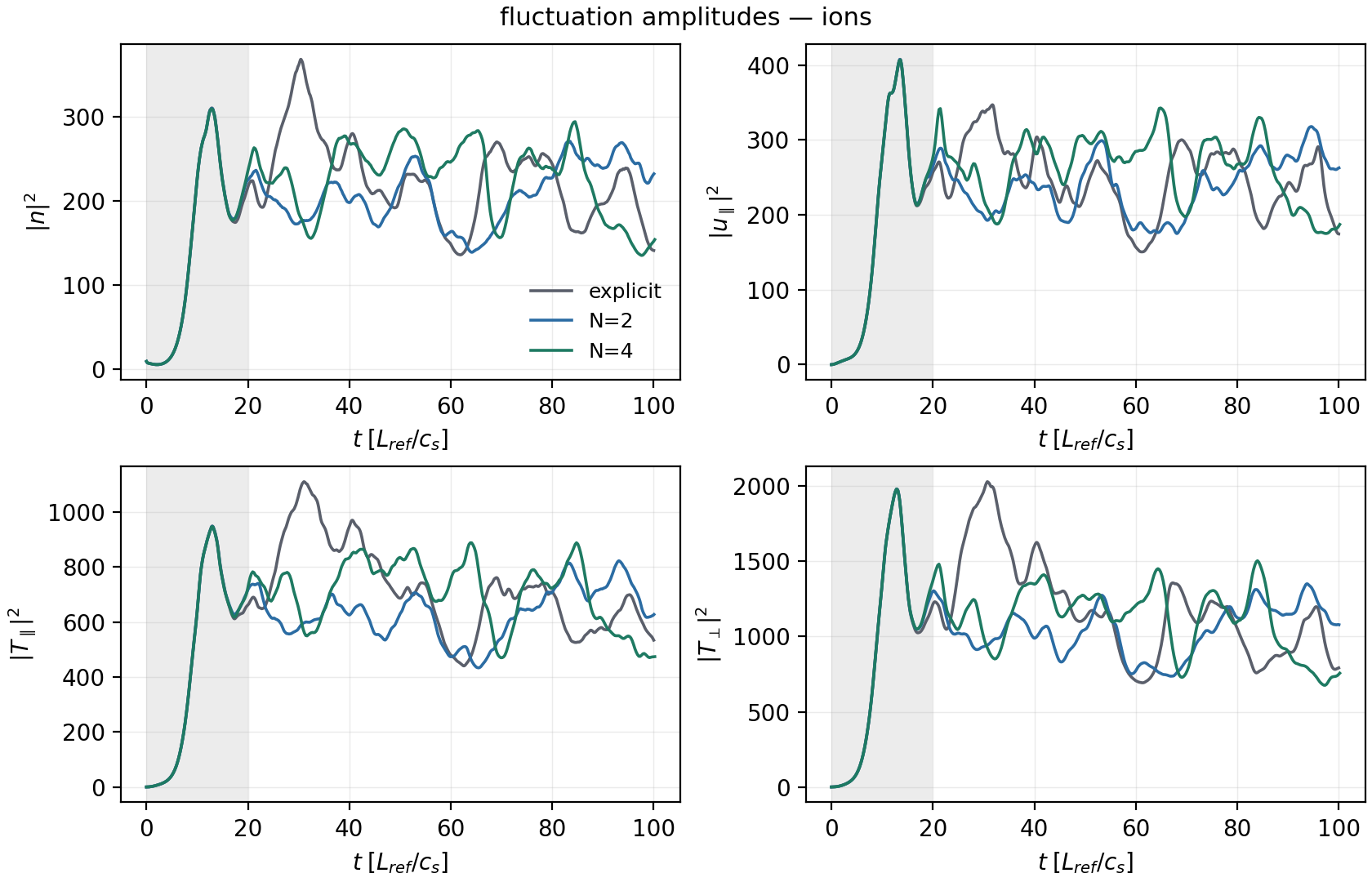}
\caption{Same as Figure \ref{fig:subcycle_fluxes} showing the behavior of fluctuations with respect to the sub-cycling rate N.}
\label{fig:subcycle_fluc}
\end{figure}

\section{Implicit parallel streaming}\label{App:implicit}
We successfully implemented an implicit treatment of parallel streaming in the electrostatic limit, reproducing the explicit results to the splitting order but we could not obtain a fast enough  implementation for the default electromagnetic cases. In this case, (neglecting $B_\parallel$ for simplicity) the implicit solve then couples $\varphi$ through quasineutrality and $A_\parallel$ through Amp\`ere, closed by a two-field response matrix for the stacked unknown $[\varphi,A_\parallel]$. The computational cost, however, is paid in the parallel direction that is often decomposed for production. The twist-and-shift boundary condition connects different $k_x$ modes into chains, and each chain must be gathered into a single extended-$z$ line before the solve. That line  spans the entire parallel domain, so under parallel decomposition it crosses  every device holding the field line: the implicit stage needs a gather-solve-scatter along each chain at every step, where the explicit scheme  needs only a nearest-neighbor exchange of ghost points. Furthermore, the response matrix grows with the extended-$z$ length, itself set by the connection number, so at the target resolution and realistic mass ratio the restarted Krylov iteration failed to converge at the small Krylov dimension that fitted a single device memory, while the larger dimension that would have converged did not fit. The stability gain never materialized as wall-clock gain on more than one device, and the scheme was thus abandoned.


\providecommand{\newblock}{}

\end{document}